\documentclass[reprint,prl,aps,nofootinbib]{revtex4-2}
\usepackage{amsfonts}
\usepackage{amsmath}
\usepackage{amssymb}
\usepackage{graphicx}
\usepackage{dcolumn}
\usepackage{bm}
\usepackage{float}
\usepackage{color}

\usepackage{pdfpages}
\makeatletter
\AtBeginDocument{\let\LS@rot\@undefined}
\makeatother

\newcommand{\func}[1]{\operatorname{#1}}

\begin{document}

\title{On-the-Fly \textit{Ab Initio} Hagedorn Wavepacket Dynamics: Single
Vibronic Level Fluorescence Spectra of Difluorocarbene}
\author{Zhan Tong Zhang}
\author{M\'at\'e Visegr\'adi}
\author{Ji\v{r}\'i J. L. Van\'i\v{c}ek}
\email{jiri.vanicek@epfl.ch}
\affiliation{Laboratory of Theoretical Physical Chemistry, Institut des Sciences et
Ing\'enierie Chimiques, Ecole Polytechnique F\'ed\'erale de Lausanne (EPFL),
CH-1015 Lausanne, Switzerland}
\date{\today}

\begin{abstract}
Hagedorn wavepackets have been used to compute single vibronic level
(SVL) spectra efficiently in model harmonic potentials. To make the
Hagedorn approach practical for realistic polyatomic molecules with
anharmonicity, here we combine local harmonic
Hagedorn wavepacket dynamics with on-the-fly \textit{ab initio} dynamics. We
then test this method by computing the SVL fluorescence spectra of
difluorocarbene, a small, floppy molecule with a very anharmonic potential
energy surface. Our time-dependent approach obtains the emission spectra of
all initial vibrational levels from a single anharmonic semiclassical
wavepacket trajectory without the need to fit individual anharmonic
vibrational wavefunctions and to calculate the Franck--Condon factors for
all vibronic transitions. We show that, whereas global harmonic models are
inadequate for CF$_2$, the spectra computed with the on-the-fly local
harmonic Hagedorn wavepacket dynamics agree well with experimental data,
especially for low initial excitations.
\end{abstract}

\maketitle

\graphicspath{{./swp/Figures/}{./Figures/}{"C:/Users/GROUP
LCPT/Documents/Tong/SVL_LHA_cf2/Figures/"}}

In the single vibronic level (SVL) fluorescence experiment, a molecule is
first excited by a precise light source to a specific vibrational level in
the electronic excited state, and its emission from that level is then
measured. This emission spectrum provides information on the excited-state
relaxation processes and on the higher vibrational levels of the ground
electronic state~\cite{Schlag_Weyssenhoff:1969,Parmenter_Schuyler:1970,Felker_Zewail:1985,Quack_Stockburger:1972,Smith_Clouthier:2022a,Suzuki_Okuyama:2024}.

Following Tapavicza's generating function approach~\cite{Tapavicza:2019} and
our derivation of algebraic expressions for the overlaps between any two
Hagedorn functions~\cite{Vanicek_Zhang:2024}, we recently developed a
simple, time-dependent method for computing SVL spectra with Hagedorn
wavepacket dynamics~\cite{Zhang_Vanicek:2024a,Zhang_Vanicek:2024b,Zhang_Vanicek:2024c}. In this
method, instead of individually evaluating the Franck--Condon overlaps for
all vibronic transitions, the final spectrum is computed from the
autocorrelation function between the initial wavepacket and the wavepacket
propagated on the final electronic surface. This approach avoids the need to
preselect the relevant transitions and can account for mode-mixing
(Duschinsky rotation) and anharmonic effects more efficiently, avoiding the
calculations of peaks that are not resolved in the experimental spectra. In
addition, the Hagedorn approach has the advantage that a single Gaussian
wavepacket trajectory is sufficient to compute the SVL emission spectra from
any initial vibronic level.

Our approach was validated in harmonic model potentials~\cite{Zhang_Vanicek:2024a}, where the Hagedorn wavepackets are exact solutions to
the time-dependent Schr\"{o}dinger equation, and it was successfully applied
to anthracene using a global harmonic model constructed from density
functional theory calculations~\cite{Zhang_Vanicek:2024b}. To investigate
whether the Hagedorn method can capture moderate anharmonicity, we combined
it with the local harmonic approximation (LHA) and, in a proof-of-principle
study, demonstrated an improvement over global harmonic models in Morse
systems for which exact quantum calculations were possible~\cite{Zhang_Vanicek:2024c}.

%Furthermore, we have combined it with the local harmonic approximation (LHA) to capture effects from anharmonicity in SVL spectra and demonstrated the improvement offered by LHA over global harmonic models in Morse-type model potentials.~\cite{Zhang_Vanicek:2024c}

To make the method practical for real polyatomic molecules, here we build on
these preliminary results by combining the local harmonic Hagedorn
wavepacket approach with on-the-fly \textit{ab initio} semiclassical~\cite{Miller:2001}
dynamics~\cite{Tatchen_Pollak:2009,Ceotto_Aspuru-Guzik:2009,Wong_Roy:2011,Saita_Shalashilin:2012,Ceotto_Conte:2017,DiLiberto_Ceotto:2018,Pios_Chen:2024}.
Similar to the thawed Gaussian approximation~\cite{Heller:1975,book_Heller:2018} 
used for evaluating ground-level
emission and absorption spectra~\cite{Wehrle_Vanicek:2014,Wehrle_Vanicek:2015,Begusic_Vanicek:2022,Kletnieks_Vanicek:2023,Gherib_Genin:2024},
the trajectory-based Hagedorn wavepacket dynamics avoids the need for precomputing a full anharmonic potential energy surface and instead relies on
potential energy information evaluated locally during propagation. To test
the on-the-fly Hagedorn approach on a realistic system, we compute the SVL
spectra of difluorocarbene (CF$_{2}$), for which experimental results are
available from the ground level up to the sixth excitation in the bending
mode~\cite{King_Stephenson:1979}. By extending the LHA to Hagedorn
wavepacket dynamics, we aim to compute the emission from higher vibrational
levels and simultaneously include the anharmonic effects without the need
for variational or perturbative correction schemes. 
%We obtain good agreement with experimental results, and while the deviation in higher vibrational levels becomes greater, the local harmonic Hagedorn approach still performs reasonably, especially for describing the splitting of the spectral envelope due to the initial vibrational excitation.

In the Hagedorn wavepacket approach to SVL spectroscopy~\cite{Zhang_Vanicek:2024a}, the initial state $|K\rangle \equiv |e, K\rangle$
with vibrational quantum numbers $K = (K_1,\dots,K_D) \in \mathbb{N}^D_0$ in
the electronic excited state $e$ is represented in the excited-state
normal-mode coordinates by the Hagedorn function~\cite{Hagedorn:1981,Hagedorn:1985},
\begin{equation}
\varphi_{K} = (K!)^{-1/2} (A^{\dagger})^{K} \varphi_0,  \label{eq:hgf}
\end{equation}
constructed from a $D$-dimensional, normalized, complex-valued Gaussian
wavepacket (representing the ground vibrational wavefunction), 
\begin{multline}
\varphi_{0}(q) = \frac{1}{(\pi \hbar)^{D / 4} \sqrt{\det (Q_{t}) }} \\
\times \exp \left\{ \frac{i}{\hbar} \left[ \frac{1}{2} x^{T} \cdot P_{t}
\cdot Q_{t}^{- 1} \cdot x + p_{t}^{T} \cdot x + S_{t} \right] \right\},
\label{eq:tga}
\end{multline}
by applying the Hagedorn raising operator~\cite{Hagedorn:1998,Lasser_Lubich:2020},
\begin{equation}
A^{\dagger} := \frac{i}{\sqrt{2 \hbar}} \left(P_{t}^{\dagger} \cdot (\hat{q}
- q_{t}) - Q_{t}^{\dagger} \cdot (\hat{p} - p_t) \right).  \label{eq:raising}
\end{equation}
In Eq.~(\ref{eq:hgf}), $K! = K_{1}!\cdot K_{2}! \cdots  K_{D}!$, and $%
(A^{\dagger})^{K}=(A^{\dagger}_{1})^{K_{1}}\cdot
(A^{\dagger}_{2})^{K_{2}}\cdots(A^{\dagger}_{D})^{K_{D}}$, where $%
A^{\dagger}_{j}$ is the $j$-th component of the raising operator $A^{\dagger}
$. In Eq.~(\ref{eq:tga}), $x:= q - q_{t}$ is the shifted position, and the
Gaussian wavepacket $\varphi_{0}$ is parametrized by its ``classical''
position $q_{t}$ and momentum $p_{t}$, two complex-valued $D$-dimensional
matrices $Q_{t}$ and $P_{t}$ (such that $P_{t}\cdot Q_{t}^{-1}$ is the
complex symmetric width matrix), and the classical action $S_t$. Matrices $%
Q_{t}$ and $P_{t}$, related to the position and momentum covariances of the
Gaussian~\cite{Vanicek:2023}, satisfy the so-called symplecticity conditions given by Eqs.~(2.1) and (2.2) of Ref.~\cite{Faou_Lubich:2009}, which ensure the commutation relations
$[A_j, A_k]=[A_j^{\dagger}, A_k^{\dagger}]=0$ and $ [A_j, A_k^{\dagger}]=\delta_{j k}$
among Hagedorn's raising operators $A_{j}^{\dagger}$ and lowering operators $A_{k}$.
%To satisfy the conditions~(\ref{eqn:symp_rel1}) and (\ref{eqn:symp_rel2})
%\begin{align}
%Q_{t}^{T} \cdot P_{t} - P_{t}^{T} \cdot Q_{t} &= 0,  \label{eqn:symp_rel1} \\
%Q_{t}^{\dagger} \cdot P_{t} - P_{t}^{\dagger} \cdot Q_{t} &
%= 2 i \mathrm{Id},  \label{eqn:symp_rel2}
%\end{align}
%which ensure the commutator relations
%\begin{align}
%    [A_j, A_k^{\dagger}]=\delta_{j k}, \qquad [A_j, A_k]=[A_j^{\dagger}, A_k^{\dagger}]=0
%\end{align}
%between the different components of the raising operator $A^{\dagger}$ and the corresponding lowering operator $A$~\cite{Hagedorn:1998,book_Lubich:2008}.
Parametrization (\ref{eq:tga}), originally proposed by Hagedorn~\cite{Hagedorn:1980,Faou_Lubich:2009},
facilitates the construction of Hagedorn functions (\ref{eq:hgf}) from Eqs.~(\ref{eq:tga}) and (\ref{eq:raising}). These functions have the form of a Gaussian multiplied
by a polynomial~\cite{Hagedorn:1998,Lasser_Lubich:2020} and are related to
the generalized coherent states~\cite{Combescure:1992,Combescure:2012,Lasser_Troppmann:2014,Borrelli_Gelin:2016a,Chen_Zhao:2017}.

To meet the symplecticity conditions, 
we choose the initial parameters $Q_0 = (\operatorname{Im} \mathrm{A}_0)^{-1/2}$ and $P_0 = \mathrm{A}_0 \cdot Q_0$, where $\mathrm{A_0}$ is the width matrix of the ground vibrational Gaussian wavefunction in the excited electronic state.
By propagating the initial vibrational wavepacket $|K\rangle =
|\varphi_{K}\rangle$ on the electronic ground-state surface $V_{g}$, we
obtain the autocorrelation function 
\begin{equation}
C(t)= \langle \varphi_{K} | e^{-i H_{g}t/\hbar} | \varphi_{K} \rangle,
\label{eqn:autocorr}
\end{equation}
whose Fourier transform~\cite{Heller:1981a,book_Tannor:2007,Tapavicza:2019} 
\begin{equation}
\sigma_{\text{em}}(\omega) = \frac{4\omega^3|{\mu} _{ge}|^2}{3\pi \hbar c^3} 
\func{Re} \int^\infty_{0} \overline{C(t)}\,e^{it(\omega -
\omega_{e,K})} dt  \label{eq:spec_ft}
\end{equation}
gives the rate of spontaneous emission from the vibrational level $K$ in the
excited electronic state $e$ to the ground electronic state $g$ as a
function of frequency $\omega$. In Eqs.~(\ref{eqn:autocorr})
and (\ref{eq:spec_ft}), $H_g$ is the ground-state Hamiltonian, $\mu_{ge}$ is the
(constant) transition dipole moment within the Condon approximation,
and $\hbar\omega_{e,K}$ is the vibronic energy of the initial state.

In the local harmonic approximation, the ground-state potential $V_g$ is
expanded at each time $t$ around the current center $q_t$ of the wavepacket
as 
\begin{equation}
V_{\text{LHA}}(q; q_t) := V(q_{t}) + V^{\prime}(q_{t})\cdot x + x^{T}\cdot
V^{\prime\prime}(q_{t}) \cdot x/2.
\end{equation}
Remarkably, Hagedorn wavepackets remain exact solutions to the
time-dependent Schr\"{o}dinger equation under this local harmonic potential $%
V_{\text{LHA}}$, with a set of particularly simple, classical-like
equations, 
\begin{align}
\dot{q_t} &= m^{-1} \cdot p_t, & \dot{p_t} &= -V^{\prime}(q_t)  \notag \\
\dot{Q_t} &= m^{-1} \cdot P_t, & \dot{P_t} &= -V^{\prime\prime}(q_t) \cdot
Q_t,  \notag \\
\dot{S_t} &= L_t, &   \label{eq:eom}
\end{align}
where $m$ is the mass matrix and $L_{t}$ is the Lagrangian~\cite{book_Lubich:2008,Faou_Lubich:2009,Lasser_Lubich:2020,Vanicek:2023}. These equations of
motion are identical to those in the thawed Gaussian approximation, and the
local harmonic propagation of the Hagedorn functions depends simply on the
evolution of the Gaussian. Therefore, the SVL spectra from \emph{all}
vibrational levels may be computed ``for free" from the \emph{same}
trajectory of the five parameters $q_t, p_t, Q_t, P_t, S_t$, needed already
to evaluate the ground-level emission. The overlaps between the two Hagedorn
wavepackets in the autocorrelation function (\ref{eqn:autocorr}) can be
computed from the recurrence relations derived in Ref.~%
\cite{Vanicek_Zhang:2024}. The cost of these overlaps is negligible
compared with the cost of \textit{ab initio} calculations required to
propagate the guiding Gaussian.

We assume that the initial wavepacket is a vibrationally excited state of a
harmonic potential, which can be represented either by a single Hagedorn
function (\ref{eq:hgf}) with a diagonal $Q_{0,jj} = {(m_j\omega_j)^{-1/2}}$
matrix or, equivalently, by the product 
\begin{equation}
\varphi_{K}(q) = \langle q|K\rangle = \frac{ \varphi_0(q)}{\sqrt{2^{|K|}K!}}
\prod_{j=1}^D H_{K_j}\left( \sqrt{\frac{m\omega_j}{\hbar}}\cdot x_j\right)
\label{eq:hag_vibr}
\end{equation}
of a Gaussian and $D$ univariate Hermite polynomials~\cite{Lasser_Lubich:2020,Zhang_Vanicek:2024a}. Here, we used the mass-weighted
normal-mode coordinates of the initial, excited electronic surface, and $%
\omega_j$ is the angular frequency of the $j$-th vibrational mode with a
common mass $m_{j} = m$. In general, however, Hagedorn functions are not
simple products of one-dimensional Hermite functions~\cite{Lasser_Troppmann:2014,Ohsawa:2019}. Under the effect of Duschinsky
rotation, the excited-state vibrational eigenfunction will not remain
separable if it is propagated with the ground-state potential. Yet, with the
local harmonic Hagedorn wavepacket dynamics, the multi-index $K$ of each
Hagedorn function $\varphi_K$ remains constant during the propagation of the
parameters of the associated Gaussian.

The accuracy of the local harmonic dynamics of Hagedorn function $\varphi_K$ depends on the anharmonicity of the potential $V$. Hagedorn~\cite{Hagedorn:1998} showed analytically that the $L^2$-norm of the propagated wavefunction is of order $\mathcal{O}(\hbar^{1/2})$ under certain technical assumptions on the potential.
In the one-dimensional case, he also proved that the error bound grows with the function's excitation as $K^{3/2}$ and depends on $V^{\prime\prime\prime}(q)$. Alternatively, the accuracy of our approach can be studied numerically by simply comparing the local harmonic and exact quantum spectra; this was done for Morse-type model potentials in Ref.~\cite{Zhang_Vanicek:2024c}. When the exact potential energy surface is known but exact quantum calculation is impractical, a measure of accuracy may be derived by integrating the difference between the true and local harmonic Hamiltonians over time
~\cite{Lasser_Lubich:2020,Burkhard_Lasser:2024} (thus estimating the ``uncaptured" anharmonicity), similar to how the deviation vector is used to assess the variational dynamics of the Davydov ansatz~\cite{Zhao_Gelin:2022}.

%In practical molecular applications, the quality of electronic structure calculations also contributes to the error relative to experimental spectra. Where neither a global surface nor experimental data are available, the adequacy of the local harmonic Hagedorn approach could be assessed by evaluating higher-order derivatives of the potential with an accurate electronic structure method.

\begin{figure*}[!hbt]
\centering
\includegraphics[width=0.85\linewidth]{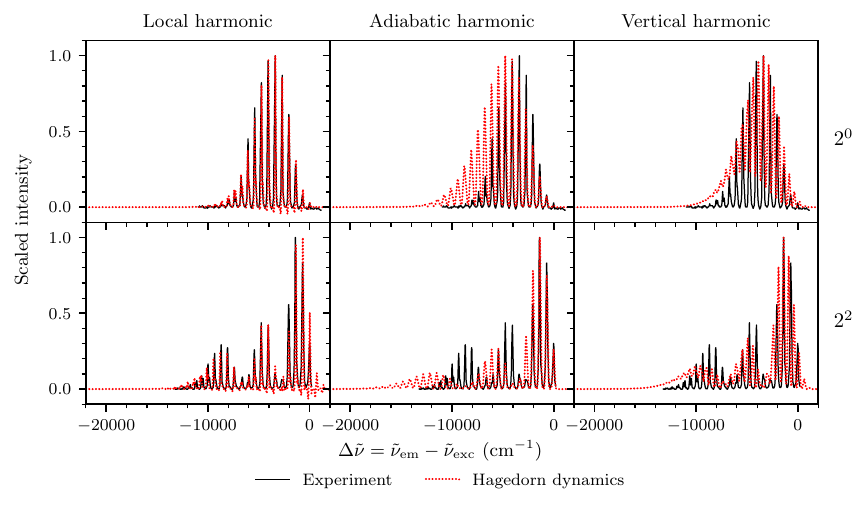}  
\caption{Comparison of the experimental (black solid line) SVL fluorescence
spectra from $2^0$ and $2^2$ levels of CF$_2$ from Ref.~
\cite{King_Stephenson:1979} with computed (red dotted line) spectra
using the local, adiabatic, and vertical harmonic Hagedorn wavepacket
dynamics; the intensity is plotted as a function of the difference $\Delta\tilde{\nu} = \tilde{\nu}_\text{em} - \tilde{\nu}_\text{exc}$ between the wavenumbers of the emitted photon and of the light used for the initial vibronic excitation.}
\label{fig:lha_vs_gha}
\end{figure*}

To assess the accuracy of the local harmonic Hagedorn wavepacket dynamics in
a realistic system, here we compute the emission spectra from the vibrational
ground ($2^0$) level and the first six excited vibrational levels ($2^1$ to $%
2^6$) in the bending mode ($v_2$) of difluorocarbene (CF$_2$) and compare
them to the experiments~\cite{King_Stephenson:1979}.

Difluorocarbene plays an important role in organic synthesis and is formed during plasma etching and chlorofluorocarbon decomposition processes~\cite{Ma_Song:2023,Xie_Hu:2024,Bulcourt_Dyke:2004,Cuddy_Fisher:2012,Rebbert_Ausloos:1975,Sonoyama_Sakata:2002}. With only
three vibrational degrees of freedom, it also serves as a simple yet
realistic model to test methods for simulating spectra beyond the global
harmonic approximation. 
%As a reactive intermediate, CF$_2$ was the subject of extensive spectroscopic studies~\cite{Milligan_Mitsch:1964,Powell_Lide:1966,Mathews_Mathews:1967,King_Stephenson:1979,Millard_Kay:1982,Cameron_Bacskay:1995,Cameron_Kable:1998,Innocenti_Dyke:2008} and was used to explore plasma processes~\cite{Bulcourt_Dyke:2004,Liu_Fisher:2009,Huebner_Helden:2015} and chemical reactions.~\cite{Cobos_Troe:2017,Cobos_Troe:2021}
%Being a flexible molecule, difluorocarbene has a potential energy surface with significant anharmonicity, the effects of which are particularly important in spectral calculations. 

Chau et al.~\cite{Chau_Mok:2001} simulated the $\tilde{A}^1 B_1 \rightarrow 
\tilde{X}^1 {A}_1$ SVL spectra of CF$_{2}$ by fitting the potential energy
function in a special form, variationally determining the anharmonic
vibrational wavefunctions as a sum of products of one-dimensional harmonic
oscillator eigenfunctions, and subsequently iteratively computing the
Franck--Condon factors for each transition using the time-independent
sum-over-states expression of vibronic spectroscopy. 
%It was found that including anharmonic corrections was necessary to reproduce the experimental results accurately, particularly for fluorescence spectra from higher vibrational levels.

In our time-dependent method, the local harmonic SVL spectra were evaluated
from a single anharmonic on-the-fly wavepacket trajectory, obtained by
evaluating the potential energy $V$, gradient $V^{\prime}$, and Hessian $%
V^{\prime\prime}$ locally at each time step and propagating the wavepacket
according to Eqs.~(\ref{eq:eom}). All electronic structure calculations were
performed with Gaussian 16~\cite{software_g16} at the PBE0/aug-cc-pVTZ level
of theory~\cite{Adamo_Barone:1999,Kendall_Harrison:1992}. See the
Supplemental Material for the optimized geometries and frequencies of CF$_2$
in the ground and first excited singlet states compared to the experimental
data.

To investigate the importance of anharmonicity, we simulated the $2^0$ and $%
2^2$ spectra also using Hagedorn dynamics combined with one of two global
harmonic approximations~\cite{Zhang_Vanicek:2024a,Zhang_Vanicek:2024b}. The
adiabatic harmonic model employs the expansion 
\begin{equation}
V_{\text{AHA}}(q) = V_{g}(q_{\text{eq},g}) +(q-q_{\text{eq},g})^T\cdot
V^{\prime\prime}_{g}(q_{\text{ref},g}) \cdot (q-q_{\text{eq},g})/2 
\label{eq:aha}
\end{equation}
of the potential $V_{g}$ around the optimized ground-state equilibrium
geometry $q_{\text{eq},g}$, whereas the vertical harmonic model employs the
expansion 
\begin{multline}
V_{\text{VHA}}(q) = V_{g}(q_{\text{eq},e}) + V^{\prime}_{g}(q_{\text{eq},e})^T
\cdot (q-q_{\text{eq},e}) \\
+ (q-q_{\text{eq},e})^T\cdot V^{\prime\prime}_{g}(q_{\text{eq},e}) \cdot
(q-q_{\text{eq},e})/2  \label{eq:vha}
\end{multline}
of the potential around the Franck--Condon point, that is, the optimized
excited-state equilibrium geometry $q_{\text{eq},e}$.

\begin{figure*}[!t]
\centering
\includegraphics[width=0.685\linewidth]{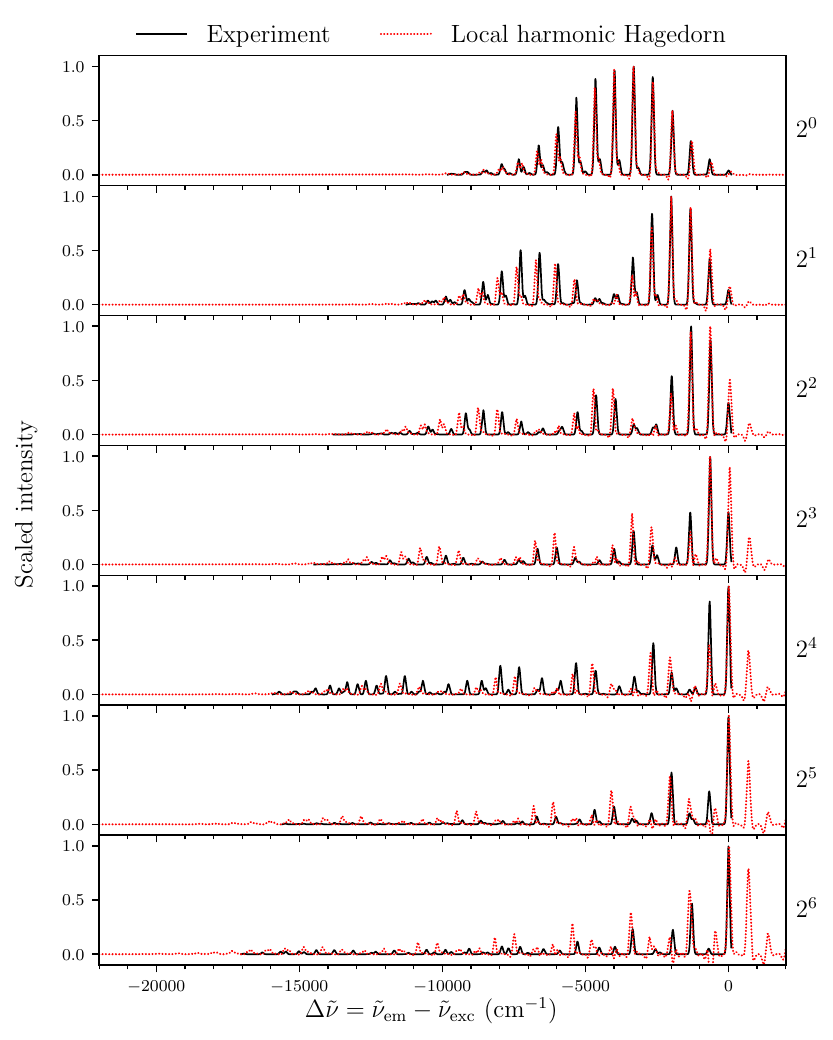}  
\caption{Comparison of computed (red dotted line) and experimental (black
solid line) fluorescence spectra of CF$_2$ from $2^0$ to $2^6$ levels, plotted as a function of the difference $\Delta\tilde\nu = \tilde{\nu}_\text{em}-\tilde{\nu}_\text{exc}$ between the wavenumbers of the emitted photon and of the initial excitation light; the
experimental spectra are artificially broadened based on wavelength and
intensity values from Table I of Ref.~\cite{King_Stephenson:1979}. }
\label{fig:lha_expt}
\end{figure*}

In both the local and global harmonic simulations, the parameters of the
Gaussian were propagated for a total time of $16000\,\text{a.u.}$ ($\sim$%
387\,fs) using the second-order TVT geometric integrator~\cite{book_Lubich:2008,Vanicek:2023} based on the Strang splitting~\cite{book_Hairer_Wanner:2006}.
A small time step of 8\,a.u. was used to ensure the
accuracy of the dynamics. 
For each initial vibrational level, the autocorrelation function was computed every two steps. As the spectrum (\ref{eq:spec_ft}) is the Fourier transform
of the autocorrelation function, calculating the wavepacket overlaps less frequently (i.e., at every other step) reduces only the spectral range, which remains sufficient to cover the relevant part of the spectra in our case (see Figs.~\ref{fig:lha_vs_gha} and \ref{fig:lha_expt}).
The overlaps between the Hagedorn wavepackets were evaluated using the algebraic algorithm derived in Ref.~%
\cite{Vanicek_Zhang:2024} (see Eqs.~97 and 98 therein) and the publicly available code from
Ref.~\cite{Zhang_Vanicek:2024a}. 
A Gaussian damping function was applied to
the autocorrelation functions such that the spectral peaks were broadened by
a Gaussian with a half-width at half-maximum of $50\,\text{cm}^{-1}$.

Figure~\ref{fig:lha_vs_gha} compares the emission spectra from vibrational
levels $2^0$ and $2^2$ computed using the local and global harmonic models
with the experimental spectra. The wavenumbers of experimental
spectra were shifted such that the transition to the vibrational ground level in each  spectrum was at $0\,%
\text{cm}^{-1}$. In other words, the intensity is plotted as a function of
the difference $\Delta\tilde{\nu} = \tilde{\nu}_\text{em} - \tilde{\nu}_\text{exc}$ between the wavenumbers of
the emitted light and of the initial excitation light;  the peaks at more negative wavenumbers correspond to transitions to higher vibrational levels in the ground electronic state. Before the $\omega^3$ scaling factor in Eq.~(\ref{eq:spec_ft}) was applied,
each computed spectrum was horizontally shifted to remove the relatively large error in the electronic excitation energy provided by the electronic structure program (see the Supplementary Material for details).
To facilitate the comparison, we scaled both
the computed and experimental spectra by setting the intensity of the
highest peak in each spectrum to unity.
%Whereas the experimental spectra were scaled by the peak with the largest intensity,  the computed spectra were heuristically scaled to improve the agreement with experimental data since we are most interested in the relative intensity and peak positions.
%In \cref{fig:lha_vs_gha}, the spectra obtained in the vertical harmonic model were scaled by their highest peaks. For the local and adiabatic harmonic cases, the computed $2^0$ spectra were scaled by the peak with the largest intensity, whereas the computed $2^2$ spectra were scaled so that the $2^2_2$ peaks (indicated by arrows) have the same intensity as the experiment.

In the ground-level spectrum ($2^0$, first row), the adiabatic harmonic
model describes the peaks closer to $0\,\text{cm}^{-1}$ reasonably well and provides the
correct spacing between the peaks. However, it significantly overestimates
the intensity of the transitions to the higher vibrational levels ($\Delta\tilde{\nu}<\!-5000\,\text{cm}^{-1}$) in the ground state.
The vertical harmonic model captures the overall envelope of
the spectra quite well but yields a poorly resolved spectrum with a broad tail at more negative values of $\Delta\tilde{\nu}$.

The delocalization of the wavepacket increases with the initial vibrational
excitation and amplifies the effects of anharmonicity. Although the
adiabatic model qualitatively captures the splitting of the envelope in the $%
2^2$ spectrum (second row), the splitting positions and the intensity
patterns are incorrect. The spectrum of the vertical harmonic model also
exhibits splitting of the envelope and arguably captures it better than the
adiabatic model, but the resolution remains poor at lower wavenumbers.

In contrast to the global harmonic models, the local harmonic dynamics
(first column of Fig.~\ref{fig:lha_vs_gha}) reproduces remarkably well the
experimental peak positions and intensities in both $2^0$ and $2^2$ spectra. 
%Despite overestimating the peaks for $\Delta\tilde{\nu}<\!-8000\,\text{cm}^{-1}$,
In the $2^2$ case (second row), the local harmonic spectrum predicts correctly
the locations of the envelope splitting. The significant improvement
observed with the local harmonic Hagedorn wavepacket dynamics highlights the
importance of including anharmonic effects in the SVL spectra of small,
floppy molecules, particularly with a higher initial excitation, even when
the ground-level spectrum computed with global harmonic models is
satisfactory.

Next, we employed the same on-the-fly \textit{ab initio} trajectory that we
had used for the $2^0$ and $2^2$ spectra to compute the emission spectra of CF$_2$
also from levels $2^1, 2^3, 2^4, 2^5$ and $2^6$.
Each of these SVL spectra is dominated by the bending-mode (mode 2) progression, whose envelope is split according to the initial vibrational excitation~\cite{King_Stephenson:1979}.
To facilitate the
comparison of the computed and experimental spectra in Fig.~\ref{fig:lha_expt},
we applied artificial broadening to the experimental values~\cite{King_Stephenson:1979} of the wavelengths
and spectral intensities of the emission peaks from $2^0$ to $2^6$ levels.

In Fig.~\ref{fig:lha_expt}, the local harmonic spectra deviate more from the
experimental ones as the initial vibrational excitation increases. However,
the local harmonic dynamics still captures the correct splitting pattern due
to the vibrational excitation of the initial state and yields reasonable
peak spacing and intensities for $\Delta\tilde{\nu} > -5000\,\text{cm}^{-1}$ even with relatively high initial vibrational excitations.

Although the same \textit{ab initio} trajectory is used, the performance of our approach becomes poorer for spectra with higher initial vibrational excitations, which is in agreement with Hagedorn's error bound~\cite{Hagedorn:1998}.
Even if the initial, excited-state surface is completely harmonic, an initial nuclear wavepacket with higher vibrational excitation becomes more delocalized and, when propagated on the anharmonic ground-state surface, is exposed to stronger anharmonic effects
in regions farther from the wavepacket's center, where the local harmonic approximation becomes insufficient.
In addition, the assumption of a harmonic excited-state surface, which justifies
representing the initial wavepacket by a single Hagedorn function, may break
down as the initial vibrational excitation increases.

Nonetheless, the local harmonic Hagedorn wavepacket approach is suitable for
evaluating SVL spectra with relatively low excitations, and its on-the-fly
implementation is very efficient. The emission spectra from all vibrational
levels are obtained from a single trajectory without the need to rerun
expensive electronic structure calculations for different initial states.
This advantage becomes even more important in higher-dimensional systems containing more
atoms and electrons.
Additionally, the local harmonic approach may potentially perform better in larger
molecules that are less floppy and hence less anharmonic than CF$_2$.

Alternative approaches employing time-dependent bases, such as the 
multiconfigurational time-dependent Hartree method~\cite{Meyer_Cederbaum:1990,Beck_Meyer:2000},
multiple spawning~\cite{Ben-Nun_Martinez:2000},
coupled coherent states~\cite{Shalashilin_Child:2004a},
variational multiconfigurational Gaussians~\cite{Worth_Burghardt:2004},
Gauss--Hermite basis~\cite{Adhikari_Billing:1998,Billing:2002},
and multiple Davydov ansatz~\cite{Zhou_Zhao:2015,Zhao_Gelin:2022,Zhao:2023},
have been used to obtain highly accurate results for molecular quantum dynamics.
Similarly, our approach can be made more accurate by expanding the wavepacket as a linear combination of multiple Hagedorn functions (associated with the same guiding Gaussian)
and propagating the expansion coefficients using the time-dependent variational principle~\cite{Faou_Lubich:2009,Lasser_Lubich:2020}. Such variational approach has been applied to several model problems~\cite{Faou_Lubich:2009,Kieri_Karlsson:2012,Zhou:2014,Gradinaru_Rietmann:2021},
and its errors and other properties have been rigorously studied by
mathematicians~\cite{Hagedorn:1980,Hagedorn_Joye:2000,book_Lubich:2008,Gradinaru_Hagedorn:2014,Ohsawa:2018,Lasser_Lubich:2020}.

Our local harmonic Hagedorn approach, however, offers a quick and straightforward way to incorporate anharmonic effects in vibronic spectra arising from vibrationally excited states, using almost the same mathematical machinery as in the global harmonic case~\cite{Zhang_Vanicek:2024a,Zhang_Vanicek:2024b}.
In contrast, the more accurate variational Hagedorn method requires a separate, variational propagation of the coefficients for each initial excitation and, as dimensions grow, an increasing number of basis functions. An on-the-fly variational implementation would also require efficient electronic structure methods capable of providing accurate third- or higher-order derivatives.

In conclusion, we successfully combined the local harmonic Hagedorn
wavepacket approach with on-the-fly \textit{ab initio} molecular dynamics to
simulate the SVL fluorescence spectra of polyatomic molecules. Using a
single semiclassical wavepacket trajectory, we were able to evaluate SVL
emission spectra from different initial vibronic levels of the small and
flexible CF$_2$ molecule. Whereas the global harmonic models proved to be
inadequate for CF$_2$, the local harmonic Hagedorn method resulted in good
agreement with the experimental data, at least for lower initial vibrational
excitations. Propagation with the local harmonic approximation requires
potential energy surface information evaluated only at a single molecular
geometry at each time step, and obtaining SVL emission spectra from
arbitrary initial vibrational levels incurs minimal additional computational
expenses with Hagedorn dynamics. Thus, our efficient method can be applied
to larger molecules, similar to how the thawed Gaussian approximation has
been used for emission and absorption spectra from the ground vibrational
level~\cite{Wehrle_Vanicek:2015,Begusic_Vanicek:2019,Begusic_Vanicek:2020}.
Finally, the on-the-fly local harmonic Hagedorn wavepacket approach should
be applicable to
other vibronic spectroscopy techniques involving excited
vibrational states~\cite{VanWilderen_Bredenbeck:2014,Yu_Ullrich:2014,Baiardi_Barone:2014,WhaleyMayda_Tokmakoff:2021,
Lau_Neumark:2023,Horz_Burghardt:2023} and to more situations with non-Gaussian initial states~\cite{Lee_Heller:1982, Tannor_Heller:1982},
such as the computation of Herzberg--Teller spectra~\cite{Baiardi_Barone:2013,Bonfanti_Pollak:2018,Patoz_Vanicek:2018,Kundu_Makri:2022}
and internal conversion rates~\cite{Wenzel_Mitric:2023}.

\begin{acknowledgments}
The authors acknowledge financial support from the EPFL.
\end{acknowledgments}

\bibliographystyle{apsrev4-2}

\begin{thebibliography}{86}%
\makeatletter
\providecommand \@ifxundefined [1]{%
 \@ifx{#1\undefined}
}%
\providecommand \@ifnum [1]{%
 \ifnum #1\expandafter \@firstoftwo
 \else \expandafter \@secondoftwo
 \fi
}%
\providecommand \@ifx [1]{%
 \ifx #1\expandafter \@firstoftwo
 \else \expandafter \@secondoftwo
 \fi
}%
\providecommand \natexlab [1]{#1}%
\providecommand \enquote  [1]{``#1''}%
\providecommand \bibnamefont  [1]{#1}%
\providecommand \bibfnamefont [1]{#1}%
\providecommand \citenamefont [1]{#1}%
\providecommand \href@noop [0]{\@secondoftwo}%
\providecommand \href [0]{\begingroup \@sanitize@url \@href}%
\providecommand \@href[1]{\@@startlink{#1}\@@href}%
\providecommand \@@href[1]{\endgroup#1\@@endlink}%
\providecommand \@sanitize@url [0]{\catcode `\\12\catcode `\$12\catcode `\&12\catcode `\#12\catcode `\^12\catcode `\_12\catcode `\%12\relax}%
\providecommand \@@startlink[1]{}%
\providecommand \@@endlink[0]{}%
\providecommand \url  [0]{\begingroup\@sanitize@url \@url }%
\providecommand \@url [1]{\endgroup\@href {#1}{\urlprefix }}%
\providecommand \urlprefix  [0]{URL }%
\providecommand \Eprint [0]{\href }%
\providecommand \doibase [0]{https://doi.org/}%
\providecommand \selectlanguage [0]{\@gobble}%
\providecommand \bibinfo  [0]{\@secondoftwo}%
\providecommand \bibfield  [0]{\@secondoftwo}%
\providecommand \translation [1]{[#1]}%
\providecommand \BibitemOpen [0]{}%
\providecommand \bibitemStop [0]{}%
\providecommand \bibitemNoStop [0]{.\EOS\space}%
\providecommand \EOS [0]{\spacefactor3000\relax}%
\providecommand \BibitemShut  [1]{\csname bibitem#1\endcsname}%
\let\auto@bib@innerbib\@empty
%</preamble>
\bibitem [{\citenamefont {Schlag}\ and\ \citenamefont {v.~Weyssenhoff}(1969)}]{Schlag_Weyssenhoff:1969}%
  \BibitemOpen
  \bibfield  {author} {\bibinfo {author} {\bibfnamefont {E.~W.}\ \bibnamefont {Schlag}}\ and\ \bibinfo {author} {\bibfnamefont {H.}~\bibnamefont {v.~Weyssenhoff}},\ }\href {https://doi.org/10.1063/1.1672373} {\bibfield  {journal} {\bibinfo  {journal} {J.~Chem.\ Phys.}\ }\textbf {\bibinfo {volume} {51}},\ \bibinfo {pages} {2508} (\bibinfo {year} {1969})}\BibitemShut {NoStop}%
\bibitem [{\citenamefont {Parmenter}\ and\ \citenamefont {Schuyler}(1970)}]{Parmenter_Schuyler:1970}%
  \BibitemOpen
  \bibfield  {author} {\bibinfo {author} {\bibfnamefont {C.}~\bibnamefont {Parmenter}}\ and\ \bibinfo {author} {\bibfnamefont {M.}~\bibnamefont {Schuyler}},\ }in\ \href@noop {} {\emph {\bibinfo {booktitle} {Transitions non radiatives dans les molécules : 20e réunion de la Société de chimie physique, Paris, 27 au 30 mai 1969}}}\ (\bibinfo {organization} {Société de chimie physique},\ \bibinfo {year} {1970})\ p.~\bibinfo {pages} {92}\BibitemShut {NoStop}%
\bibitem [{\citenamefont {Felker}\ and\ \citenamefont {Zewail}(1985)}]{Felker_Zewail:1985}%
  \BibitemOpen
  \bibfield  {author} {\bibinfo {author} {\bibfnamefont {P.~M.}\ \bibnamefont {Felker}}\ and\ \bibinfo {author} {\bibfnamefont {A.~H.}\ \bibnamefont {Zewail}},\ }\href {https://doi.org/10.1063/1.448247} {\bibfield  {journal} {\bibinfo  {journal} {J.~Chem.\ Phys.}\ }\textbf {\bibinfo {volume} {82}},\ \bibinfo {pages} {2975} (\bibinfo {year} {1985})}\BibitemShut {NoStop}%
\bibitem [{\citenamefont {Quack}\ and\ \citenamefont {Stockburger}(1972)}]{Quack_Stockburger:1972}%
  \BibitemOpen
  \bibfield  {author} {\bibinfo {author} {\bibfnamefont {M.}~\bibnamefont {Quack}}\ and\ \bibinfo {author} {\bibfnamefont {M.}~\bibnamefont {Stockburger}},\ }\href {https://doi.org/10.1016/0022-2852(72)90164-6} {\bibfield  {journal} {\bibinfo  {journal} {J.~Mol.\ Spec.}\ }\textbf {\bibinfo {volume} {43}},\ \bibinfo {pages} {87} (\bibinfo {year} {1972})}\BibitemShut {NoStop}%
\bibitem [{\citenamefont {Smith}\ \emph {et~al.}(2022)\citenamefont {Smith}, \citenamefont {Gharaibeh},\ and\ \citenamefont {Clouthier}}]{Smith_Clouthier:2022a}%
  \BibitemOpen
  \bibfield  {author} {\bibinfo {author} {\bibfnamefont {T.~C.}\ \bibnamefont {Smith}}, \bibinfo {author} {\bibfnamefont {M.}~\bibnamefont {Gharaibeh}},\ and\ \bibinfo {author} {\bibfnamefont {D.~J.}\ \bibnamefont {Clouthier}},\ }\href {https://doi.org/10.1063/5.0127449} {\bibfield  {journal} {\bibinfo  {journal} {J.~Chem.\ Phys.}\ }\textbf {\bibinfo {volume} {157}},\ \bibinfo {pages} {204306} (\bibinfo {year} {2022})}\BibitemShut {NoStop}%
\bibitem [{\citenamefont {Suzuki}\ \emph {et~al.}(2024)\citenamefont {Suzuki}, \citenamefont {Chiba}, \citenamefont {Tanaka},\ and\ \citenamefont {Okuyama}}]{Suzuki_Okuyama:2024}%
  \BibitemOpen
  \bibfield  {author} {\bibinfo {author} {\bibfnamefont {R.}~\bibnamefont {Suzuki}}, \bibinfo {author} {\bibfnamefont {K.}~\bibnamefont {Chiba}}, \bibinfo {author} {\bibfnamefont {S.}~\bibnamefont {Tanaka}},\ and\ \bibinfo {author} {\bibfnamefont {K.}~\bibnamefont {Okuyama}},\ }\href {https://doi.org/10.1063/5.0176162} {\bibfield  {journal} {\bibinfo  {journal} {J.~Chem.\ Phys.}\ }\textbf {\bibinfo {volume} {160}},\ \bibinfo {pages} {024301} (\bibinfo {year} {2024})}\BibitemShut {NoStop}%
\bibitem [{\citenamefont {Tapavicza}(2019)}]{Tapavicza:2019}%
  \BibitemOpen
  \bibfield  {author} {\bibinfo {author} {\bibfnamefont {E.}~\bibnamefont {Tapavicza}},\ }\href {https://doi.org/10.1021/acs.jpclett.9b02273} {\bibfield  {journal} {\bibinfo  {journal} {J.~Phys.\ Chem.\ Lett.}\ }\textbf {\bibinfo {volume} {10}},\ \bibinfo {pages} {6003} (\bibinfo {year} {2019})}\BibitemShut {NoStop}%
\bibitem [{\citenamefont {Van\'{\i}\v{c}ek}\ and\ \citenamefont {Zhang}(2024)}]{Vanicek_Zhang:2024}%
  \BibitemOpen
  \bibfield  {author} {\bibinfo {author} {\bibfnamefont {J.~J.~L.}\ \bibnamefont {Van\'{\i}\v{c}ek}}\ and\ \bibinfo {author} {\bibfnamefont {Z.~T.}\ \bibnamefont {Zhang}},\ }\href {https://arxiv.org/abs/2405.07880} {\bibinfo {title} {On {Hagedorn} wavepackets associated with different {Gaussians}}} (\bibinfo {year} {2024}),\ \bibinfo {note} {arXiv:2405.07880 [quant-ph]}\BibitemShut {NoStop}%
\bibitem [{\citenamefont {Zhang}\ and\ \citenamefont {Van\'{\i}\v{c}ek}(2024{\natexlab{a}})}]{Zhang_Vanicek:2024a}%
  \BibitemOpen
  \bibfield  {author} {\bibinfo {author} {\bibfnamefont {Z.~T.}\ \bibnamefont {Zhang}}\ and\ \bibinfo {author} {\bibfnamefont {J.~J.~L.}\ \bibnamefont {Van\'{\i}\v{c}ek}},\ }\href {https://doi.org/doi:10.1063/5.0219005} {\bibfield  {journal} {\bibinfo  {journal} {J.~Chem.\ Phys.}\ }\textbf {\bibinfo {volume} {161}},\ \bibinfo {pages} {111101} (\bibinfo {year} {2024}{\natexlab{a}})}\BibitemShut {NoStop}%
\bibitem [{\citenamefont {Zhang}\ and\ \citenamefont {Van\'{\i}\v{c}ek}(2024{\natexlab{b}})}]{Zhang_Vanicek:2024b}%
  \BibitemOpen
  \bibfield  {author} {\bibinfo {author} {\bibfnamefont {Z.~T.}\ \bibnamefont {Zhang}}\ and\ \bibinfo {author} {\bibfnamefont {J.~J.~L.}\ \bibnamefont {Van\'{\i}\v{c}ek}},\ }\href {https://arxiv.org/abs/2403.00702} {\bibinfo {title} {Ab initio simulation of single vibronic level fluorescence spectra of anthracene using {H}agedorn wavepackets}} (\bibinfo {year} {2024}{\natexlab{b}}),\ \bibinfo {note} {arXiv:2403.00702 [physics.chem-ph]}\BibitemShut {NoStop}%
\bibitem [{\citenamefont {Zhang}\ \emph {et~al.}(2024)\citenamefont {Zhang}, \citenamefont {Visegr\'{a}di},\ and\ \citenamefont {Van\'{\i}\v{c}ek}}]{Zhang_Vanicek:2024c}%
  \BibitemOpen
  \bibfield  {author} {\bibinfo {author} {\bibfnamefont {Z.~T.}\ \bibnamefont {Zhang}}, \bibinfo {author} {\bibfnamefont {M.}~\bibnamefont {Visegr\'{a}di}},\ and\ \bibinfo {author} {\bibfnamefont {J.~J.~L.}\ \bibnamefont {Van\'{\i}\v{c}ek}},\ }\href {https://arxiv.org/abs/2408.11991} {\bibinfo {title} {Capturing anharmonic effects in single vibronic level fluorescence spectra using local harmonic {Hagedorn} wavepacket dynamics}} (\bibinfo {year} {2024}),\ \bibinfo {note} {arXiv:2408.11991 [physics.chem-ph]}\BibitemShut {NoStop}%
\bibitem [{\citenamefont {Miller}(2001)}]{Miller:2001}%
  \BibitemOpen
  \bibfield  {author} {\bibinfo {author} {\bibfnamefont {W.~H.}\ \bibnamefont {Miller}},\ }\href {https://doi.org/10.1021/jp003712k} {\bibfield  {journal} {\bibinfo  {journal} {J.~Phys.\ Chem.~A}\ }\textbf {\bibinfo {volume} {105}},\ \bibinfo {pages} {2942} (\bibinfo {year} {2001})}\BibitemShut {NoStop}%
\bibitem [{\citenamefont {Tatchen}\ and\ \citenamefont {Pollak}(2009)}]{Tatchen_Pollak:2009}%
  \BibitemOpen
  \bibfield  {author} {\bibinfo {author} {\bibfnamefont {J.}~\bibnamefont {Tatchen}}\ and\ \bibinfo {author} {\bibfnamefont {E.}~\bibnamefont {Pollak}},\ }\href {https://doi.org/10.1063/1.3074100} {\bibfield  {journal} {\bibinfo  {journal} {J.~Chem.\ Phys.}\ }\textbf {\bibinfo {volume} {130}},\ \bibinfo {pages} {041103} (\bibinfo {year} {2009})}\BibitemShut {NoStop}%
\bibitem [{\citenamefont {Ceotto}\ \emph {et~al.}(2009)\citenamefont {Ceotto}, \citenamefont {Atahan}, \citenamefont {Shim}, \citenamefont {Tantardini},\ and\ \citenamefont {Aspuru-Guzik}}]{Ceotto_Aspuru-Guzik:2009}%
  \BibitemOpen
  \bibfield  {author} {\bibinfo {author} {\bibfnamefont {M.}~\bibnamefont {Ceotto}}, \bibinfo {author} {\bibfnamefont {S.}~\bibnamefont {Atahan}}, \bibinfo {author} {\bibfnamefont {S.}~\bibnamefont {Shim}}, \bibinfo {author} {\bibfnamefont {G.~F.}\ \bibnamefont {Tantardini}},\ and\ \bibinfo {author} {\bibfnamefont {A.}~\bibnamefont {Aspuru-Guzik}},\ }\href {https://doi.org/10.1039/B820785B} {\bibfield  {journal} {\bibinfo  {journal} {Phys.\ Chem.\ Chem.\ Phys.}\ }\textbf {\bibinfo {volume} {11}},\ \bibinfo {pages} {3861} (\bibinfo {year} {2009})}\BibitemShut {NoStop}%
\bibitem [{\citenamefont {Wong}\ \emph {et~al.}(2011)\citenamefont {Wong}, \citenamefont {Benoit}, \citenamefont {Lewerenz}, \citenamefont {Brown},\ and\ \citenamefont {Roy}}]{Wong_Roy:2011}%
  \BibitemOpen
  \bibfield  {author} {\bibinfo {author} {\bibfnamefont {S.~Y.~Y.}\ \bibnamefont {Wong}}, \bibinfo {author} {\bibfnamefont {D.~M.}\ \bibnamefont {Benoit}}, \bibinfo {author} {\bibfnamefont {M.}~\bibnamefont {Lewerenz}}, \bibinfo {author} {\bibfnamefont {A.}~\bibnamefont {Brown}},\ and\ \bibinfo {author} {\bibfnamefont {P.-N.}\ \bibnamefont {Roy}},\ }\href {https://doi.org/10.1063/1.3553179} {\bibfield  {journal} {\bibinfo  {journal} {J.~Chem.\ Phys.}\ }\textbf {\bibinfo {volume} {134}},\ \bibinfo {pages} {094110} (\bibinfo {year} {2011})}\BibitemShut {NoStop}%
\bibitem [{\citenamefont {Saita}\ and\ \citenamefont {Shalashilin}(2012)}]{Saita_Shalashilin:2012}%
  \BibitemOpen
  \bibfield  {author} {\bibinfo {author} {\bibfnamefont {K.}~\bibnamefont {Saita}}\ and\ \bibinfo {author} {\bibfnamefont {D.~V.}\ \bibnamefont {Shalashilin}},\ }\href {https://doi.org/10.1063/1.4734313} {\bibfield  {journal} {\bibinfo  {journal} {J.~Chem.\ Phys.}\ }\textbf {\bibinfo {volume} {137}},\ \bibinfo {eid} {22A506} (\bibinfo {year} {2012})}\BibitemShut {NoStop}%
\bibitem [{\citenamefont {Ceotto}\ \emph {et~al.}(2017)\citenamefont {Ceotto}, \citenamefont {Di~Liberto},\ and\ \citenamefont {Conte}}]{Ceotto_Conte:2017}%
  \BibitemOpen
  \bibfield  {author} {\bibinfo {author} {\bibfnamefont {M.}~\bibnamefont {Ceotto}}, \bibinfo {author} {\bibfnamefont {G.}~\bibnamefont {Di~Liberto}},\ and\ \bibinfo {author} {\bibfnamefont {R.}~\bibnamefont {Conte}},\ }\href {https://doi.org/10.1103/PhysRevLett.119.010401} {\bibfield  {journal} {\bibinfo  {journal} {Phys.\ Rev.\ Lett.}\ }\textbf {\bibinfo {volume} {119}},\ \bibinfo {pages} {010401} (\bibinfo {year} {2017})}\BibitemShut {NoStop}%
\bibitem [{\citenamefont {Di~Liberto}\ \emph {et~al.}(2018)\citenamefont {Di~Liberto}, \citenamefont {Conte},\ and\ \citenamefont {Ceotto}}]{DiLiberto_Ceotto:2018}%
  \BibitemOpen
  \bibfield  {author} {\bibinfo {author} {\bibfnamefont {G.}~\bibnamefont {Di~Liberto}}, \bibinfo {author} {\bibfnamefont {R.}~\bibnamefont {Conte}},\ and\ \bibinfo {author} {\bibfnamefont {M.}~\bibnamefont {Ceotto}},\ }\href {https://doi.org/10.1063/1.5023155} {\bibfield  {journal} {\bibinfo  {journal} {J.~Chem.\ Phys.}\ }\textbf {\bibinfo {volume} {148}},\ \bibinfo {pages} {104302} (\bibinfo {year} {2018})}\BibitemShut {NoStop}%
\bibitem [{\citenamefont {Pios}\ \emph {et~al.}(2024)\citenamefont {Pios}, \citenamefont {Gelin}, \citenamefont {Vasquez}, \citenamefont {Hauer},\ and\ \citenamefont {Chen}}]{Pios_Chen:2024}%
  \BibitemOpen
  \bibfield  {author} {\bibinfo {author} {\bibfnamefont {S.~V.}\ \bibnamefont {Pios}}, \bibinfo {author} {\bibfnamefont {M.~F.}\ \bibnamefont {Gelin}}, \bibinfo {author} {\bibfnamefont {L.}~\bibnamefont {Vasquez}}, \bibinfo {author} {\bibfnamefont {J.}~\bibnamefont {Hauer}},\ and\ \bibinfo {author} {\bibfnamefont {L.}~\bibnamefont {Chen}},\ }\href {https://doi.org/10.1021/acs.jpclett.4c01842} {\bibfield  {journal} {\bibinfo  {journal} {J.~Phys.\ Chem.\ Lett.}\ }\textbf {\bibinfo {volume} {15}},\ \bibinfo {pages} {8728} (\bibinfo {year} {2024})}\BibitemShut {NoStop}%
\bibitem [{\citenamefont {Heller}(1975)}]{Heller:1975}%
  \BibitemOpen
  \bibfield  {author} {\bibinfo {author} {\bibfnamefont {E.~J.}\ \bibnamefont {Heller}},\ }\href {https://doi.org/10.1063/1.430620} {\bibfield  {journal} {\bibinfo  {journal} {J.~Chem.\ Phys.}\ }\textbf {\bibinfo {volume} {62}},\ \bibinfo {pages} {1544} (\bibinfo {year} {1975})}\BibitemShut {NoStop}%
\bibitem [{\citenamefont {Heller}(2018)}]{book_Heller:2018}%
  \BibitemOpen
  \bibfield  {author} {\bibinfo {author} {\bibfnamefont {E.~J.}\ \bibnamefont {Heller}},\ }\href@noop {} {\emph {\bibinfo {title} {The Semiclassical Way to Dynamics and Spectroscopy}}}\ (\bibinfo  {publisher} {Princeton University Press},\ \bibinfo {address} {Princeton, NJ},\ \bibinfo {year} {2018})\BibitemShut {NoStop}%
\bibitem [{\citenamefont {Wehrle}\ \emph {et~al.}(2014)\citenamefont {Wehrle}, \citenamefont {\v{S}ulc},\ and\ \citenamefont {Van\'{i}\v{c}ek}}]{Wehrle_Vanicek:2014}%
  \BibitemOpen
  \bibfield  {author} {\bibinfo {author} {\bibfnamefont {M.}~\bibnamefont {Wehrle}}, \bibinfo {author} {\bibfnamefont {M.}~\bibnamefont {\v{S}ulc}},\ and\ \bibinfo {author} {\bibfnamefont {J.}~\bibnamefont {Van\'{i}\v{c}ek}},\ }\href {https://doi.org/10.1063/1.4884718} {\bibfield  {journal} {\bibinfo  {journal} {J.~Chem.\ Phys.}\ }\textbf {\bibinfo {volume} {140}},\ \bibinfo {pages} {244114} (\bibinfo {year} {2014})}\BibitemShut {NoStop}%
\bibitem [{\citenamefont {Wehrle}\ \emph {et~al.}(2015)\citenamefont {Wehrle}, \citenamefont {Oberli},\ and\ \citenamefont {Van\'{i}\v{c}ek}}]{Wehrle_Vanicek:2015}%
  \BibitemOpen
  \bibfield  {author} {\bibinfo {author} {\bibfnamefont {M.}~\bibnamefont {Wehrle}}, \bibinfo {author} {\bibfnamefont {S.}~\bibnamefont {Oberli}},\ and\ \bibinfo {author} {\bibfnamefont {J.}~\bibnamefont {Van\'{i}\v{c}ek}},\ }\href {https://doi.org/10.1021/acs.jpca.5b03907} {\bibfield  {journal} {\bibinfo  {journal} {J.~Phys.\ Chem.~A}\ }\textbf {\bibinfo {volume} {119}},\ \bibinfo {pages} {5685} (\bibinfo {year} {2015})}\BibitemShut {NoStop}%
\bibitem [{\citenamefont {Begu\v{s}i\'{c}}\ \emph {et~al.}(2022)\citenamefont {Begu\v{s}i\'{c}}, \citenamefont {Tapavicza},\ and\ \citenamefont {Van\'i\v{c}ek}}]{Begusic_Vanicek:2022}%
  \BibitemOpen
  \bibfield  {author} {\bibinfo {author} {\bibfnamefont {T.}~\bibnamefont {Begu\v{s}i\'{c}}}, \bibinfo {author} {\bibfnamefont {E.}~\bibnamefont {Tapavicza}},\ and\ \bibinfo {author} {\bibfnamefont {J.}~\bibnamefont {Van\'i\v{c}ek}},\ }\href {https://doi.org/10.1021/acs.jctc.2c00030} {\bibfield  {journal} {\bibinfo  {journal} {J.~Chem.\ Theory Comput.}\ }\textbf {\bibinfo {volume} {18}},\ \bibinfo {pages} {3065} (\bibinfo {year} {2022})}\BibitemShut {NoStop}%
\bibitem [{\citenamefont {Kl\={e}tnieks}\ \emph {et~al.}(2023)\citenamefont {Kl\={e}tnieks}, \citenamefont {Alonso},\ and\ \citenamefont {Van\'i\v{c}ek}}]{Kletnieks_Vanicek:2023}%
  \BibitemOpen
  \bibfield  {author} {\bibinfo {author} {\bibfnamefont {E.}~\bibnamefont {Kl\={e}tnieks}}, \bibinfo {author} {\bibfnamefont {Y.~C.}\ \bibnamefont {Alonso}},\ and\ \bibinfo {author} {\bibfnamefont {J.~J.~L.}\ \bibnamefont {Van\'i\v{c}ek}},\ }\href {https://doi.org/10.1021/acs.jpca.3c04607} {\bibfield  {journal} {\bibinfo  {journal} {J.~Phys.\ Chem.~A}\ }\textbf {\bibinfo {volume} {127}},\ \bibinfo {pages} {8117} (\bibinfo {year} {2023})}\BibitemShut {NoStop}%
\bibitem [{\citenamefont {Gherib}\ \emph {et~al.}(2024)\citenamefont {Gherib}, \citenamefont {Ryabinkin},\ and\ \citenamefont {Genin}}]{Gherib_Genin:2024}%
  \BibitemOpen
  \bibfield  {author} {\bibinfo {author} {\bibfnamefont {R.}~\bibnamefont {Gherib}}, \bibinfo {author} {\bibfnamefont {I.~G.}\ \bibnamefont {Ryabinkin}},\ and\ \bibinfo {author} {\bibfnamefont {S.~N.}\ \bibnamefont {Genin}},\ }\href {https://arxiv.org/abs/2405.00193} {\bibinfo {title} {Thawed {Gaussian} wavepacket dynamics with {$\Delta$}-machine learned potentials}} (\bibinfo {year} {2024}),\ \bibinfo {note} {arXiv:2405.00193 [physics.chem-ph]}\BibitemShut {NoStop}%
\bibitem [{\citenamefont {King}\ \emph {et~al.}(1979)\citenamefont {King}, \citenamefont {Schenck},\ and\ \citenamefont {Stephenson}}]{King_Stephenson:1979}%
  \BibitemOpen
  \bibfield  {author} {\bibinfo {author} {\bibfnamefont {D.~S.}\ \bibnamefont {King}}, \bibinfo {author} {\bibfnamefont {P.~K.}\ \bibnamefont {Schenck}},\ and\ \bibinfo {author} {\bibfnamefont {J.~C.}\ \bibnamefont {Stephenson}},\ }\href {https://doi.org/10.1016/0022-2852(79)90031-6} {\bibfield  {journal} {\bibinfo  {journal} {J.~Mol.\ Spec.}\ }\textbf {\bibinfo {volume} {78}},\ \bibinfo {pages} {1} (\bibinfo {year} {1979})}\BibitemShut {NoStop}%
\bibitem [{\citenamefont {Hagedorn}(1981)}]{Hagedorn:1981}%
  \BibitemOpen
  \bibfield  {author} {\bibinfo {author} {\bibfnamefont {G.~A.}\ \bibnamefont {Hagedorn}},\ }\href {https://doi.org/10.1016/0003-4916(81)90143-3} {\bibfield  {journal} {\bibinfo  {journal} {Ann.\ Phys.\ (NY)}\ }\textbf {\bibinfo {volume} {135}},\ \bibinfo {pages} {58} (\bibinfo {year} {1981})}\BibitemShut {NoStop}%
\bibitem [{\citenamefont {Hagedorn}(1985)}]{Hagedorn:1985}%
  \BibitemOpen
  \bibfield  {author} {\bibinfo {author} {\bibfnamefont {G.~A.}\ \bibnamefont {Hagedorn}},\ }\href {http://www.numdam.org/item/AIHPA_1985__42_4_363_0/} {\bibfield  {journal} {\bibinfo  {journal} {Ann. Henri Poincar{\'e}}\ }\textbf {\bibinfo {volume} {42}},\ \bibinfo {pages} {363} (\bibinfo {year} {1985})}\BibitemShut {NoStop}%
\bibitem [{\citenamefont {Hagedorn}(1998)}]{Hagedorn:1998}%
  \BibitemOpen
  \bibfield  {author} {\bibinfo {author} {\bibfnamefont {G.~A.}\ \bibnamefont {Hagedorn}},\ }\href {https://doi.org/10.1006/aphy.1998.5843} {\bibfield  {journal} {\bibinfo  {journal} {Ann.\ Phys.\ (NY)}\ }\textbf {\bibinfo {volume} {269}},\ \bibinfo {pages} {77} (\bibinfo {year} {1998})}\BibitemShut {NoStop}%
\bibitem [{\citenamefont {Lasser}\ and\ \citenamefont {Lubich}(2020)}]{Lasser_Lubich:2020}%
  \BibitemOpen
  \bibfield  {author} {\bibinfo {author} {\bibfnamefont {C.}~\bibnamefont {Lasser}}\ and\ \bibinfo {author} {\bibfnamefont {C.}~\bibnamefont {Lubich}},\ }\href {https://doi.org/10.1017/S0962492920000033} {\bibfield  {journal} {\bibinfo  {journal} {Acta Numerica}\ }\textbf {\bibinfo {volume} {29}},\ \bibinfo {pages} {229} (\bibinfo {year} {2020})}\BibitemShut {NoStop}%
\bibitem [{\citenamefont {Van\'i\v{c}ek}(2023)}]{Vanicek:2023}%
  \BibitemOpen
  \bibfield  {author} {\bibinfo {author} {\bibfnamefont {J.~J.~L.}\ \bibnamefont {Van\'i\v{c}ek}},\ }\href {https://doi.org/10.1063/5.0146680} {\bibfield  {journal} {\bibinfo  {journal} {J.~Chem.\ Phys.}\ }\textbf {\bibinfo {volume} {159}},\ \bibinfo {pages} {014114} (\bibinfo {year} {2023})}\BibitemShut {NoStop}%
\bibitem [{\citenamefont {Faou}\ \emph {et~al.}(2009)\citenamefont {Faou}, \citenamefont {Gradinaru},\ and\ \citenamefont {Lubich}}]{Faou_Lubich:2009}%
  \BibitemOpen
  \bibfield  {author} {\bibinfo {author} {\bibfnamefont {E.}~\bibnamefont {Faou}}, \bibinfo {author} {\bibfnamefont {V.}~\bibnamefont {Gradinaru}},\ and\ \bibinfo {author} {\bibfnamefont {C.}~\bibnamefont {Lubich}},\ }\href {https://doi.org/10.1137/080729724} {\bibfield  {journal} {\bibinfo  {journal} {SIAM J.\ Sci.\ Comp.}\ }\textbf {\bibinfo {volume} {31}},\ \bibinfo {pages} {3027} (\bibinfo {year} {2009})}\BibitemShut {NoStop}%
\bibitem [{\citenamefont {Hagedorn}(1980)}]{Hagedorn:1980}%
  \BibitemOpen
  \bibfield  {author} {\bibinfo {author} {\bibfnamefont {G.~A.}\ \bibnamefont {Hagedorn}},\ }\href {https://doi.org/10.1007/BF01230088} {\bibfield  {journal} {\bibinfo  {journal} {Commun.\ Math.\ Phys.}\ }\textbf {\bibinfo {volume} {71}},\ \bibinfo {pages} {77} (\bibinfo {year} {1980})}\BibitemShut {NoStop}%
\bibitem [{\citenamefont {Combescure}(1992)}]{Combescure:1992}%
  \BibitemOpen
  \bibfield  {author} {\bibinfo {author} {\bibfnamefont {M.}~\bibnamefont {Combescure}},\ }\href {https://doi.org/10.1063/1.529884} {\bibfield  {journal} {\bibinfo  {journal} {J.~Math.\ Phys.}\ }\textbf {\bibinfo {volume} {33}},\ \bibinfo {pages} {3870} (\bibinfo {year} {1992})}\BibitemShut {NoStop}%
\bibitem [{\citenamefont {Combescure}\ and\ \citenamefont {Robert}(2012)}]{Combescure:2012}%
  \BibitemOpen
  \bibfield  {author} {\bibinfo {author} {\bibfnamefont {M.}~\bibnamefont {Combescure}}\ and\ \bibinfo {author} {\bibfnamefont {D.}~\bibnamefont {Robert}},\ }\href@noop {} {\emph {\bibinfo {title} {Coherent States and Applications in Mathematical Physics}}},\ Theoretical and Mathematical Physics Ser.\ (\bibinfo  {publisher} {Springer Netherlands},\ \bibinfo {address} {Dordrecht},\ \bibinfo {year} {2012})\BibitemShut {NoStop}%
\bibitem [{\citenamefont {Lasser}\ and\ \citenamefont {Troppmann}(2014)}]{Lasser_Troppmann:2014}%
  \BibitemOpen
  \bibfield  {author} {\bibinfo {author} {\bibfnamefont {C.}~\bibnamefont {Lasser}}\ and\ \bibinfo {author} {\bibfnamefont {S.}~\bibnamefont {Troppmann}},\ }\href {https://doi.org/10.1007/s00041-014-9330-9} {\bibfield  {journal} {\bibinfo  {journal} {J. Fourier Anal. Appl.}\ }\textbf {\bibinfo {volume} {20}},\ \bibinfo {pages} {679} (\bibinfo {year} {2014})}\BibitemShut {NoStop}%
\bibitem [{\citenamefont {Borrelli}\ and\ \citenamefont {Gelin}(2016)}]{Borrelli_Gelin:2016a}%
  \BibitemOpen
  \bibfield  {author} {\bibinfo {author} {\bibfnamefont {R.}~\bibnamefont {Borrelli}}\ and\ \bibinfo {author} {\bibfnamefont {M.~F.}\ \bibnamefont {Gelin}},\ }\href {https://doi.org/10.1016/j.chemphys.2016.05.013} {\bibfield  {journal} {\bibinfo  {journal} {Chem.\ Phys.}\ }\textbf {\bibinfo {volume} {481}},\ \bibinfo {pages} {91} (\bibinfo {year} {2016})}\BibitemShut {NoStop}%
\bibitem [{\citenamefont {Chen}\ \emph {et~al.}(2017)\citenamefont {Chen}, \citenamefont {Borrelli},\ and\ \citenamefont {Zhao}}]{Chen_Zhao:2017}%
  \BibitemOpen
  \bibfield  {author} {\bibinfo {author} {\bibfnamefont {L.}~\bibnamefont {Chen}}, \bibinfo {author} {\bibfnamefont {R.}~\bibnamefont {Borrelli}},\ and\ \bibinfo {author} {\bibfnamefont {Y.}~\bibnamefont {Zhao}},\ }\href {https://doi.org/10.1021/acs.jpca.7b07069} {\bibfield  {journal} {\bibinfo  {journal} {J.~Phys.\ Chem.\ A}\ }\textbf {\bibinfo {volume} {121}},\ \bibinfo {pages} {8757–8770} (\bibinfo {year} {2017})}\BibitemShut {NoStop}%
\bibitem [{\citenamefont {Heller}(1981)}]{Heller:1981a}%
  \BibitemOpen
  \bibfield  {author} {\bibinfo {author} {\bibfnamefont {E.~J.}\ \bibnamefont {Heller}},\ }\href {https://doi.org/10.1021/ar00072a002} {\bibfield  {journal} {\bibinfo  {journal} {Acc.\ Chem.\ Res.}\ }\textbf {\bibinfo {volume} {14}},\ \bibinfo {pages} {368} (\bibinfo {year} {1981})}\BibitemShut {NoStop}%
\bibitem [{\citenamefont {Tannor}(2007)}]{book_Tannor:2007}%
  \BibitemOpen
  \bibfield  {author} {\bibinfo {author} {\bibfnamefont {D.~J.}\ \bibnamefont {Tannor}},\ }\href@noop {} {\emph {\bibinfo {title} {Introduction to Quantum Mechanics: A Time-Dependent Perspective}}}\ (\bibinfo  {publisher} {University Science Books},\ \bibinfo {address} {Sausalito},\ \bibinfo {year} {2007})\BibitemShut {NoStop}%
\bibitem [{\citenamefont {Lubich}(2008)}]{book_Lubich:2008}%
  \BibitemOpen
  \bibfield  {author} {\bibinfo {author} {\bibfnamefont {C.}~\bibnamefont {Lubich}},\ }\href@noop {} {\emph {\bibinfo {title} {From Quantum to Classical Molecular Dynamics: Reduced Models and Numerical Analysis}}}\ (\bibinfo  {publisher} {European Mathematical Society},\ \bibinfo {address} {Z\"{u}rich},\ \bibinfo {year} {2008})\BibitemShut {NoStop}%
\bibitem [{\citenamefont {Ohsawa}(2019)}]{Ohsawa:2019}%
  \BibitemOpen
  \bibfield  {author} {\bibinfo {author} {\bibfnamefont {T.}~\bibnamefont {Ohsawa}},\ }\href {https://doi.org/10.1007/s00041-018-9633-3} {\bibfield  {journal} {\bibinfo  {journal} {J. Fourier Anal. Appl.}\ }\textbf {\bibinfo {volume} {25}},\ \bibinfo {pages} {1513} (\bibinfo {year} {2019})}\BibitemShut {NoStop}%
\bibitem [{\citenamefont {Burkhard}\ \emph {et~al.}(2024)\citenamefont {Burkhard}, \citenamefont {D\"{o}rich}, \citenamefont {Hochbruck},\ and\ \citenamefont {Lasser}}]{Burkhard_Lasser:2024}%
  \BibitemOpen
  \bibfield  {author} {\bibinfo {author} {\bibfnamefont {S.}~\bibnamefont {Burkhard}}, \bibinfo {author} {\bibfnamefont {B.}~\bibnamefont {D\"{o}rich}}, \bibinfo {author} {\bibfnamefont {M.}~\bibnamefont {Hochbruck}},\ and\ \bibinfo {author} {\bibfnamefont {C.}~\bibnamefont {Lasser}},\ }\href {https://doi.org/10.1088/1751-8121/ad591e} {\bibfield  {journal} {\bibinfo  {journal} {J.~Phys.~A}\ }\textbf {\bibinfo {volume} {57}},\ \bibinfo {pages} {295202} (\bibinfo {year} {2024})}\BibitemShut {NoStop}%
\bibitem [{\citenamefont {Zhao}\ \emph {et~al.}(2022)\citenamefont {Zhao}, \citenamefont {Sun}, \citenamefont {Chen},\ and\ \citenamefont {Gelin}}]{Zhao_Gelin:2022}%
  \BibitemOpen
  \bibfield  {author} {\bibinfo {author} {\bibfnamefont {Y.}~\bibnamefont {Zhao}}, \bibinfo {author} {\bibfnamefont {K.}~\bibnamefont {Sun}}, \bibinfo {author} {\bibfnamefont {L.}~\bibnamefont {Chen}},\ and\ \bibinfo {author} {\bibfnamefont {M.}~\bibnamefont {Gelin}},\ }\href {https://doi.org/10.1002/wcms.1589} {\bibfield  {journal} {\bibinfo  {journal} {WIREs Comput.\ Mol.\ Sci.}\ }\textbf {\bibinfo {volume} {12}},\ \bibinfo {pages} {e1589} (\bibinfo {year} {2022})}\BibitemShut {NoStop}%
\bibitem [{\citenamefont {Ma}\ \emph {et~al.}(2023)\citenamefont {Ma}, \citenamefont {Su},\ and\ \citenamefont {Song}}]{Ma_Song:2023}%
  \BibitemOpen
  \bibfield  {author} {\bibinfo {author} {\bibfnamefont {X.}~\bibnamefont {Ma}}, \bibinfo {author} {\bibfnamefont {J.}~\bibnamefont {Su}},\ and\ \bibinfo {author} {\bibfnamefont {Q.}~\bibnamefont {Song}},\ }\href {https://doi.org/10.1021/acs.accounts.2c00830} {\bibfield  {journal} {\bibinfo  {journal} {Acc.\ Chem.\ Res.}\ }\textbf {\bibinfo {volume} {56}},\ \bibinfo {pages} {592} (\bibinfo {year} {2023})}\BibitemShut {NoStop}%
\bibitem [{\citenamefont {Xie}\ and\ \citenamefont {Hu}(2024)}]{Xie_Hu:2024}%
  \BibitemOpen
  \bibfield  {author} {\bibinfo {author} {\bibfnamefont {Q.}~\bibnamefont {Xie}}\ and\ \bibinfo {author} {\bibfnamefont {J.}~\bibnamefont {Hu}},\ }\href {https://doi.org/10.1021/acs.accounts.3c00719} {\bibfield  {journal} {\bibinfo  {journal} {Acc.\ Chem.\ Res.}\ }\textbf {\bibinfo {volume} {57}},\ \bibinfo {pages} {693} (\bibinfo {year} {2024})}\BibitemShut {NoStop}%
\bibitem [{\citenamefont {Bulcourt}\ \emph {et~al.}(2004)\citenamefont {Bulcourt}, \citenamefont {Booth}, \citenamefont {Hudson}, \citenamefont {Luque}, \citenamefont {Mok}, \citenamefont {Lee}, \citenamefont {Chau},\ and\ \citenamefont {Dyke}}]{Bulcourt_Dyke:2004}%
  \BibitemOpen
  \bibfield  {author} {\bibinfo {author} {\bibfnamefont {N.}~\bibnamefont {Bulcourt}}, \bibinfo {author} {\bibfnamefont {J.-P.}\ \bibnamefont {Booth}}, \bibinfo {author} {\bibfnamefont {E.~A.}\ \bibnamefont {Hudson}}, \bibinfo {author} {\bibfnamefont {J.}~\bibnamefont {Luque}}, \bibinfo {author} {\bibfnamefont {D.~K.~W.}\ \bibnamefont {Mok}}, \bibinfo {author} {\bibfnamefont {E.~P.}\ \bibnamefont {Lee}}, \bibinfo {author} {\bibfnamefont {F.-T.}\ \bibnamefont {Chau}},\ and\ \bibinfo {author} {\bibfnamefont {J.~M.}\ \bibnamefont {Dyke}},\ }\href {https://doi.org/10.1063/1.1695313} {\bibfield  {journal} {\bibinfo  {journal} {J.~Chem.\ Phys.}\ }\textbf {\bibinfo {volume} {120}},\ \bibinfo {pages} {9499} (\bibinfo {year} {2004})}\BibitemShut {NoStop}%
\bibitem [{\citenamefont {Cuddy}\ and\ \citenamefont {Fisher}(2012)}]{Cuddy_Fisher:2012}%
  \BibitemOpen
  \bibfield  {author} {\bibinfo {author} {\bibfnamefont {M.~F.}\ \bibnamefont {Cuddy}}\ and\ \bibinfo {author} {\bibfnamefont {E.~R.}\ \bibnamefont {Fisher}},\ }\href {https://doi.org/10.1021/am2018546} {\bibfield  {journal} {\bibinfo  {journal} {ACS\ Appl.\ Mater.\ Interfaces}\ }\textbf {\bibinfo {volume} {4}},\ \bibinfo {pages} {1733} (\bibinfo {year} {2012})}\BibitemShut {NoStop}%
\bibitem [{\citenamefont {Rebbert}\ and\ \citenamefont {Ausloos}(1975)}]{Rebbert_Ausloos:1975}%
  \BibitemOpen
  \bibfield  {author} {\bibinfo {author} {\bibfnamefont {R.~E.}\ \bibnamefont {Rebbert}}\ and\ \bibinfo {author} {\bibfnamefont {P.~J.}\ \bibnamefont {Ausloos}},\ }\href {https://doi.org/10.1016/0047-2670(75)85023-4} {\bibfield  {journal} {\bibinfo  {journal} {J.~Photochem.}\ }\textbf {\bibinfo {volume} {4}},\ \bibinfo {pages} {419} (\bibinfo {year} {1975})}\BibitemShut {NoStop}%
\bibitem [{\citenamefont {Sonoyama}\ \emph {et~al.}(2002)\citenamefont {Sonoyama}, \citenamefont {Ezaki}, \citenamefont {Fujii},\ and\ \citenamefont {Sakata}}]{Sonoyama_Sakata:2002}%
  \BibitemOpen
  \bibfield  {author} {\bibinfo {author} {\bibfnamefont {N.}~\bibnamefont {Sonoyama}}, \bibinfo {author} {\bibfnamefont {K.}~\bibnamefont {Ezaki}}, \bibinfo {author} {\bibfnamefont {H.}~\bibnamefont {Fujii}},\ and\ \bibinfo {author} {\bibfnamefont {T.}~\bibnamefont {Sakata}},\ }\href {https://doi.org/10.1016/s0013-4686(02)00324-9} {\bibfield  {journal} {\bibinfo  {journal} {Electrochim. Acta}\ }\textbf {\bibinfo {volume} {47}},\ \bibinfo {pages} {3847} (\bibinfo {year} {2002})}\BibitemShut {NoStop}%
\bibitem [{\citenamefont {Chau}\ \emph {et~al.}(2001)\citenamefont {Chau}, \citenamefont {Dyke}, \citenamefont {Lee},\ and\ \citenamefont {Mok}}]{Chau_Mok:2001}%
  \BibitemOpen
  \bibfield  {author} {\bibinfo {author} {\bibfnamefont {F.-T.}\ \bibnamefont {Chau}}, \bibinfo {author} {\bibfnamefont {J.~M.}\ \bibnamefont {Dyke}}, \bibinfo {author} {\bibfnamefont {E.~P.~F.}\ \bibnamefont {Lee}},\ and\ \bibinfo {author} {\bibfnamefont {D.~K.~W.}\ \bibnamefont {Mok}},\ }\href {https://doi.org/10.1063/1.1398103} {\bibfield  {journal} {\bibinfo  {journal} {J.~Chem.\ Phys.}\ }\textbf {\bibinfo {volume} {115}},\ \bibinfo {pages} {5816} (\bibinfo {year} {2001})}\BibitemShut {NoStop}%
\bibitem [{\citenamefont {Frisch}\ \emph {et~al.}(2016)\citenamefont {Frisch}, \citenamefont {Trucks}, \citenamefont {Schlegel}, \citenamefont {Scuseria}, \citenamefont {Robb}, \citenamefont {Cheeseman}, \citenamefont {Scalmani}, \citenamefont {Barone}, \citenamefont {Petersson}, \citenamefont {Nakatsuji}, \citenamefont {Li}, \citenamefont {Caricato}, \citenamefont {Marenich}, \citenamefont {Bloino}, \citenamefont {Janesko}, \citenamefont {Gomperts}, \citenamefont {Mennucci}, \citenamefont {Hratchian}, \citenamefont {Ortiz}, \citenamefont {Izmaylov}, \citenamefont {Sonnenberg}, \citenamefont {Williams-Young}, \citenamefont {Ding}, \citenamefont {Lipparini}, \citenamefont {Egidi}, \citenamefont {Goings}, \citenamefont {Peng}, \citenamefont {Petrone}, \citenamefont {Henderson}, \citenamefont {Ranasinghe}, \citenamefont {Zakrzewski}, \citenamefont {Gao}, \citenamefont {Rega}, \citenamefont {Zheng}, \citenamefont {Liang}, \citenamefont {Hada}, \citenamefont {Ehara}, \citenamefont {Toyota}, \citenamefont {Fukuda},
  \citenamefont {Hasegawa}, \citenamefont {Ishida}, \citenamefont {Nakajima}, \citenamefont {Honda}, \citenamefont {Kitao}, \citenamefont {Nakai}, \citenamefont {Vreven}, \citenamefont {Throssell}, \citenamefont {Montgomery}, \citenamefont {Peralta}, \citenamefont {Ogliaro}, \citenamefont {Bearpark}, \citenamefont {Heyd}, \citenamefont {Brothers}, \citenamefont {Kudin}, \citenamefont {Staroverov}, \citenamefont {Keith}, \citenamefont {Kobayashi}, \citenamefont {Normand}, \citenamefont {Raghavachari}, \citenamefont {Rendell}, \citenamefont {Burant}, \citenamefont {Iyengar}, \citenamefont {Tomasi}, \citenamefont {Cossi}, \citenamefont {Millam}, \citenamefont {Klene}, \citenamefont {Adamo}, \citenamefont {Cammi}, \citenamefont {Ochterski}, \citenamefont {Martin}, \citenamefont {Morokuma}, \citenamefont {Farkas}, \citenamefont {Foresman},\ and\ \citenamefont {Fox}}]{software_g16}%
  \BibitemOpen
  \bibfield  {author} {\bibinfo {author} {\bibfnamefont {M.~J.}\ \bibnamefont {Frisch}}, \bibinfo {author} {\bibfnamefont {G.~W.}\ \bibnamefont {Trucks}}, \bibinfo {author} {\bibfnamefont {H.~B.}\ \bibnamefont {Schlegel}}, \bibinfo {author} {\bibfnamefont {G.~E.}\ \bibnamefont {Scuseria}}, \bibinfo {author} {\bibfnamefont {M.~A.}\ \bibnamefont {Robb}}, \bibinfo {author} {\bibfnamefont {J.~R.}\ \bibnamefont {Cheeseman}}, \bibinfo {author} {\bibfnamefont {G.}~\bibnamefont {Scalmani}}, \bibinfo {author} {\bibfnamefont {V.}~\bibnamefont {Barone}}, \bibinfo {author} {\bibfnamefont {G.~A.}\ \bibnamefont {Petersson}}, \bibinfo {author} {\bibfnamefont {H.}~\bibnamefont {Nakatsuji}}, \bibinfo {author} {\bibfnamefont {X.}~\bibnamefont {Li}}, \bibinfo {author} {\bibfnamefont {M.}~\bibnamefont {Caricato}}, \bibinfo {author} {\bibfnamefont {A.~V.}\ \bibnamefont {Marenich}}, \bibinfo {author} {\bibfnamefont {J.}~\bibnamefont {Bloino}}, \bibinfo {author} {\bibfnamefont {B.~G.}\ \bibnamefont {Janesko}}, \bibinfo {author}
  {\bibfnamefont {R.}~\bibnamefont {Gomperts}}, \bibinfo {author} {\bibfnamefont {B.}~\bibnamefont {Mennucci}}, \bibinfo {author} {\bibfnamefont {H.~P.}\ \bibnamefont {Hratchian}}, \bibinfo {author} {\bibfnamefont {J.~V.}\ \bibnamefont {Ortiz}}, \bibinfo {author} {\bibfnamefont {A.~F.}\ \bibnamefont {Izmaylov}}, \bibinfo {author} {\bibfnamefont {J.~L.}\ \bibnamefont {Sonnenberg}}, \bibinfo {author} {\bibfnamefont {D.}~\bibnamefont {Williams-Young}}, \bibinfo {author} {\bibfnamefont {F.}~\bibnamefont {Ding}}, \bibinfo {author} {\bibfnamefont {F.}~\bibnamefont {Lipparini}}, \bibinfo {author} {\bibfnamefont {F.}~\bibnamefont {Egidi}}, \bibinfo {author} {\bibfnamefont {J.}~\bibnamefont {Goings}}, \bibinfo {author} {\bibfnamefont {B.}~\bibnamefont {Peng}}, \bibinfo {author} {\bibfnamefont {A.}~\bibnamefont {Petrone}}, \bibinfo {author} {\bibfnamefont {T.}~\bibnamefont {Henderson}}, \bibinfo {author} {\bibfnamefont {D.}~\bibnamefont {Ranasinghe}}, \bibinfo {author} {\bibfnamefont {V.~G.}\ \bibnamefont
  {Zakrzewski}}, \bibinfo {author} {\bibfnamefont {J.}~\bibnamefont {Gao}}, \bibinfo {author} {\bibfnamefont {N.}~\bibnamefont {Rega}}, \bibinfo {author} {\bibfnamefont {G.}~\bibnamefont {Zheng}}, \bibinfo {author} {\bibfnamefont {W.}~\bibnamefont {Liang}}, \bibinfo {author} {\bibfnamefont {M.}~\bibnamefont {Hada}}, \bibinfo {author} {\bibfnamefont {M.}~\bibnamefont {Ehara}}, \bibinfo {author} {\bibfnamefont {K.}~\bibnamefont {Toyota}}, \bibinfo {author} {\bibfnamefont {R.}~\bibnamefont {Fukuda}}, \bibinfo {author} {\bibfnamefont {J.}~\bibnamefont {Hasegawa}}, \bibinfo {author} {\bibfnamefont {M.}~\bibnamefont {Ishida}}, \bibinfo {author} {\bibfnamefont {T.}~\bibnamefont {Nakajima}}, \bibinfo {author} {\bibfnamefont {Y.}~\bibnamefont {Honda}}, \bibinfo {author} {\bibfnamefont {O.}~\bibnamefont {Kitao}}, \bibinfo {author} {\bibfnamefont {H.}~\bibnamefont {Nakai}}, \bibinfo {author} {\bibfnamefont {T.}~\bibnamefont {Vreven}}, \bibinfo {author} {\bibfnamefont {K.}~\bibnamefont {Throssell}}, \bibinfo {author}
  {\bibfnamefont {J.~A.}\ \bibnamefont {Montgomery}, \bibfnamefont {{Jr.}}}, \bibinfo {author} {\bibfnamefont {J.~E.}\ \bibnamefont {Peralta}}, \bibinfo {author} {\bibfnamefont {F.}~\bibnamefont {Ogliaro}}, \bibinfo {author} {\bibfnamefont {M.~J.}\ \bibnamefont {Bearpark}}, \bibinfo {author} {\bibfnamefont {J.~J.}\ \bibnamefont {Heyd}}, \bibinfo {author} {\bibfnamefont {E.~N.}\ \bibnamefont {Brothers}}, \bibinfo {author} {\bibfnamefont {K.~N.}\ \bibnamefont {Kudin}}, \bibinfo {author} {\bibfnamefont {V.~N.}\ \bibnamefont {Staroverov}}, \bibinfo {author} {\bibfnamefont {T.~A.}\ \bibnamefont {Keith}}, \bibinfo {author} {\bibfnamefont {R.}~\bibnamefont {Kobayashi}}, \bibinfo {author} {\bibfnamefont {J.}~\bibnamefont {Normand}}, \bibinfo {author} {\bibfnamefont {K.}~\bibnamefont {Raghavachari}}, \bibinfo {author} {\bibfnamefont {A.~P.}\ \bibnamefont {Rendell}}, \bibinfo {author} {\bibfnamefont {J.~C.}\ \bibnamefont {Burant}}, \bibinfo {author} {\bibfnamefont {S.~S.}\ \bibnamefont {Iyengar}}, \bibinfo {author}
  {\bibfnamefont {J.}~\bibnamefont {Tomasi}}, \bibinfo {author} {\bibfnamefont {M.}~\bibnamefont {Cossi}}, \bibinfo {author} {\bibfnamefont {J.~M.}\ \bibnamefont {Millam}}, \bibinfo {author} {\bibfnamefont {M.}~\bibnamefont {Klene}}, \bibinfo {author} {\bibfnamefont {C.}~\bibnamefont {Adamo}}, \bibinfo {author} {\bibfnamefont {R.}~\bibnamefont {Cammi}}, \bibinfo {author} {\bibfnamefont {J.~W.}\ \bibnamefont {Ochterski}}, \bibinfo {author} {\bibfnamefont {R.~L.}\ \bibnamefont {Martin}}, \bibinfo {author} {\bibfnamefont {K.}~\bibnamefont {Morokuma}}, \bibinfo {author} {\bibfnamefont {O.}~\bibnamefont {Farkas}}, \bibinfo {author} {\bibfnamefont {J.~B.}\ \bibnamefont {Foresman}},\ and\ \bibinfo {author} {\bibfnamefont {D.~J.}\ \bibnamefont {Fox}},\ }\href@noop {} {\bibinfo {title} {Gaussian~16 {R}evision {C}.01}} (\bibinfo {year} {2016}),\ \bibinfo {note} {{G}aussian Inc. Wallingford CT}\BibitemShut {NoStop}%
\bibitem [{\citenamefont {Adamo}\ and\ \citenamefont {Barone}(1999)}]{Adamo_Barone:1999}%
  \BibitemOpen
  \bibfield  {author} {\bibinfo {author} {\bibfnamefont {C.}~\bibnamefont {Adamo}}\ and\ \bibinfo {author} {\bibfnamefont {V.}~\bibnamefont {Barone}},\ }\href {https://doi.org/10.1063/1.478522} {\bibfield  {journal} {\bibinfo  {journal} {J.~Chem.\ Phys.}\ }\textbf {\bibinfo {volume} {110}},\ \bibinfo {pages} {6158} (\bibinfo {year} {1999})}\BibitemShut {NoStop}%
\bibitem [{\citenamefont {Kendall}\ \emph {et~al.}(1992)\citenamefont {Kendall}, \citenamefont {Dunning},\ and\ \citenamefont {Harrison}}]{Kendall_Harrison:1992}%
  \BibitemOpen
  \bibfield  {author} {\bibinfo {author} {\bibfnamefont {R.~A.}\ \bibnamefont {Kendall}}, \bibinfo {author} {\bibfnamefont {T.~H.}\ \bibnamefont {Dunning}},\ and\ \bibinfo {author} {\bibfnamefont {R.~J.}\ \bibnamefont {Harrison}},\ }\href {https://doi.org/10.1063/1.462569} {\bibfield  {journal} {\bibinfo  {journal} {J.~Chem.\ Phys.}\ }\textbf {\bibinfo {volume} {96}},\ \bibinfo {pages} {6796} (\bibinfo {year} {1992})}\BibitemShut {NoStop}%
\bibitem [{\citenamefont {Hairer}\ \emph {et~al.}(2006)\citenamefont {Hairer}, \citenamefont {Lubich},\ and\ \citenamefont {Wanner}}]{book_Hairer_Wanner:2006}%
  \BibitemOpen
  \bibfield  {author} {\bibinfo {author} {\bibfnamefont {E.}~\bibnamefont {Hairer}}, \bibinfo {author} {\bibfnamefont {C.}~\bibnamefont {Lubich}},\ and\ \bibinfo {author} {\bibfnamefont {G.}~\bibnamefont {Wanner}},\ }\href {http://books.google.ch/books/about/Geometric_Numerical_Integration.html?id=T1TaNRLmZv8C&redir_esc=y} {\emph {\bibinfo {title} {Geometric Numerical Integration: Structure-Preserving Algorithms for Ordinary Differential Equations}}}\ (\bibinfo  {publisher} {Springer Berlin Heidelberg New York},\ \bibinfo {year} {2006})\BibitemShut {NoStop}%
\bibitem [{\citenamefont {Meyer}\ \emph {et~al.}(1990)\citenamefont {Meyer}, \citenamefont {Manthe},\ and\ \citenamefont {Cederbaum}}]{Meyer_Cederbaum:1990}%
  \BibitemOpen
  \bibfield  {author} {\bibinfo {author} {\bibfnamefont {H.-D.}\ \bibnamefont {Meyer}}, \bibinfo {author} {\bibfnamefont {U.}~\bibnamefont {Manthe}},\ and\ \bibinfo {author} {\bibfnamefont {L.~S.}\ \bibnamefont {Cederbaum}},\ }\href {https://doi.org/10.1016/0009-2614(90)87014-I} {\bibfield  {journal} {\bibinfo  {journal} {Chem.\ Phys.\ Lett.}\ }\textbf {\bibinfo {volume} {165}},\ \bibinfo {pages} {73} (\bibinfo {year} {1990})}\BibitemShut {NoStop}%
\bibitem [{\citenamefont {Beck}\ \emph {et~al.}(2000)\citenamefont {Beck}, \citenamefont {J\"{a}ckle}, \citenamefont {Worth},\ and\ \citenamefont {Meyer}}]{Beck_Meyer:2000}%
  \BibitemOpen
  \bibfield  {author} {\bibinfo {author} {\bibfnamefont {M.}~\bibnamefont {Beck}}, \bibinfo {author} {\bibfnamefont {A.}~\bibnamefont {J\"{a}ckle}}, \bibinfo {author} {\bibfnamefont {G.}~\bibnamefont {Worth}},\ and\ \bibinfo {author} {\bibfnamefont {H.-D.}\ \bibnamefont {Meyer}},\ }\href {https://doi.org/10.1016/S0370-1573(99)00047-2} {\bibfield  {journal} {\bibinfo  {journal} {Phys.\ Rep.}\ }\textbf {\bibinfo {volume} {324}},\ \bibinfo {pages} {1} (\bibinfo {year} {2000})}\BibitemShut {NoStop}%
\bibitem [{\citenamefont {Ben-Nun}\ \emph {et~al.}(2000)\citenamefont {Ben-Nun}, \citenamefont {Quenneville},\ and\ \citenamefont {Mart\'{\i}nez}}]{Ben-Nun_Martinez:2000}%
  \BibitemOpen
  \bibfield  {author} {\bibinfo {author} {\bibfnamefont {M.}~\bibnamefont {Ben-Nun}}, \bibinfo {author} {\bibfnamefont {J.}~\bibnamefont {Quenneville}},\ and\ \bibinfo {author} {\bibfnamefont {T.~J.}\ \bibnamefont {Mart\'{\i}nez}},\ }\href {https://doi.org/10.1021/jp994174i} {\bibfield  {journal} {\bibinfo  {journal} {J.~Phys.\ Chem.~A}\ }\textbf {\bibinfo {volume} {104}},\ \bibinfo {pages} {5161} (\bibinfo {year} {2000})}\BibitemShut {NoStop}%
\bibitem [{\citenamefont {Shalashilin}\ and\ \citenamefont {Child}(2004)}]{Shalashilin_Child:2004a}%
  \BibitemOpen
  \bibfield  {author} {\bibinfo {author} {\bibfnamefont {D.~V.}\ \bibnamefont {Shalashilin}}\ and\ \bibinfo {author} {\bibfnamefont {M.~S.}\ \bibnamefont {Child}},\ }\href {https://doi.org/10.1016/j.chemphys.2004.06.013} {\bibfield  {journal} {\bibinfo  {journal} {Chem.\ Phys.}\ }\textbf {\bibinfo {volume} {304}},\ \bibinfo {pages} {103} (\bibinfo {year} {2004})}\BibitemShut {NoStop}%
\bibitem [{\citenamefont {Worth}\ \emph {et~al.}(2004)\citenamefont {Worth}, \citenamefont {Robb},\ and\ \citenamefont {Burghardt}}]{Worth_Burghardt:2004}%
  \BibitemOpen
  \bibfield  {author} {\bibinfo {author} {\bibfnamefont {G.~A.}\ \bibnamefont {Worth}}, \bibinfo {author} {\bibfnamefont {M.~A.}\ \bibnamefont {Robb}},\ and\ \bibinfo {author} {\bibfnamefont {I.}~\bibnamefont {Burghardt}},\ }\href {https://doi.org/10.1039/B314253A} {\bibfield  {journal} {\bibinfo  {journal} {Faraday Discuss.}\ }\textbf {\bibinfo {volume} {127}},\ \bibinfo {pages} {307} (\bibinfo {year} {2004})}\BibitemShut {NoStop}%
\bibitem [{\citenamefont {Adhikari}\ and\ \citenamefont {Billing}(1998)}]{Adhikari_Billing:1998}%
  \BibitemOpen
  \bibfield  {author} {\bibinfo {author} {\bibfnamefont {S.}~\bibnamefont {Adhikari}}\ and\ \bibinfo {author} {\bibfnamefont {G.~D.}\ \bibnamefont {Billing}},\ }\href {https://doi.org/10.1016/s0301-0104(98)00302-4} {\bibfield  {journal} {\bibinfo  {journal} {Chem.\ Phys.}\ }\textbf {\bibinfo {volume} {238}},\ \bibinfo {pages} {69} (\bibinfo {year} {1998})}\BibitemShut {NoStop}%
\bibitem [{\citenamefont {Billing}(2002)}]{Billing:2002}%
  \BibitemOpen
  \bibfield  {author} {\bibinfo {author} {\bibfnamefont {G.~D.}\ \bibnamefont {Billing}},\ }\href {https://doi.org/10.1039/b202151j} {\bibfield  {journal} {\bibinfo  {journal} {Phys.\ Chem.\ Chem.\ Phys.}\ }\textbf {\bibinfo {volume} {4}},\ \bibinfo {pages} {2865} (\bibinfo {year} {2002})}\BibitemShut {NoStop}%
\bibitem [{\citenamefont {Zhou}\ \emph {et~al.}(2015)\citenamefont {Zhou}, \citenamefont {Huang}, \citenamefont {Zhu}, \citenamefont {Chernyak},\ and\ \citenamefont {Zhao}}]{Zhou_Zhao:2015}%
  \BibitemOpen
  \bibfield  {author} {\bibinfo {author} {\bibfnamefont {N.}~\bibnamefont {Zhou}}, \bibinfo {author} {\bibfnamefont {Z.}~\bibnamefont {Huang}}, \bibinfo {author} {\bibfnamefont {J.}~\bibnamefont {Zhu}}, \bibinfo {author} {\bibfnamefont {V.}~\bibnamefont {Chernyak}},\ and\ \bibinfo {author} {\bibfnamefont {Y.}~\bibnamefont {Zhao}},\ }\href {https://doi.org/10.1063/1.4923009} {\bibfield  {journal} {\bibinfo  {journal} {J.~Chem.\ Phys.}\ }\textbf {\bibinfo {volume} {143}},\ \bibinfo {pages} {014113} (\bibinfo {year} {2015})}\BibitemShut {NoStop}%
\bibitem [{\citenamefont {Zhao}(2023)}]{Zhao:2023}%
  \BibitemOpen
  \bibfield  {author} {\bibinfo {author} {\bibfnamefont {Y.}~\bibnamefont {Zhao}},\ }\href {https://doi.org/10.1063/5.0140002} {\bibfield  {journal} {\bibinfo  {journal} {J.~Chem.\ Phys.}\ }\textbf {\bibinfo {volume} {158}},\ \bibinfo {pages} {080901} (\bibinfo {year} {2023})}\BibitemShut {NoStop}%
\bibitem [{\citenamefont {Kieri}\ \emph {et~al.}(2012)\citenamefont {Kieri}, \citenamefont {Holmgren},\ and\ \citenamefont {Karlsson}}]{Kieri_Karlsson:2012}%
  \BibitemOpen
  \bibfield  {author} {\bibinfo {author} {\bibfnamefont {E.}~\bibnamefont {Kieri}}, \bibinfo {author} {\bibfnamefont {S.}~\bibnamefont {Holmgren}},\ and\ \bibinfo {author} {\bibfnamefont {H.~O.}\ \bibnamefont {Karlsson}},\ }\href {https://doi.org/10.1063/1.4737893} {\bibfield  {journal} {\bibinfo  {journal} {J.~Chem.\ Phys.}\ }\textbf {\bibinfo {volume} {137}},\ \bibinfo {pages} {044111} (\bibinfo {year} {2012})}\BibitemShut {NoStop}%
\bibitem [{\citenamefont {Zhou}(2014)}]{Zhou:2014}%
  \BibitemOpen
  \bibfield  {author} {\bibinfo {author} {\bibfnamefont {Z.}~\bibnamefont {Zhou}},\ }\href {https://doi.org/10.1016/j.jcp.2014.04.041} {\bibfield  {journal} {\bibinfo  {journal} {J. Comput. Phys.}\ }\textbf {\bibinfo {volume} {272}},\ \bibinfo {pages} {386} (\bibinfo {year} {2014})}\BibitemShut {NoStop}%
\bibitem [{\citenamefont {Gradinaru}\ and\ \citenamefont {Rietmann}(2021)}]{Gradinaru_Rietmann:2021}%
  \BibitemOpen
  \bibfield  {author} {\bibinfo {author} {\bibfnamefont {V.}~\bibnamefont {Gradinaru}}\ and\ \bibinfo {author} {\bibfnamefont {O.}~\bibnamefont {Rietmann}},\ }\href {https://doi.org/10.1016/j.jcp.2021.110581} {\bibfield  {journal} {\bibinfo  {journal} {J.~Comput. Phys.}\ }\textbf {\bibinfo {volume} {445}},\ \bibinfo {pages} {110581} (\bibinfo {year} {2021})}\BibitemShut {NoStop}%
\bibitem [{\citenamefont {Hagedorn}\ and\ \citenamefont {Joye}(2000)}]{Hagedorn_Joye:2000}%
  \BibitemOpen
  \bibfield  {author} {\bibinfo {author} {\bibfnamefont {G.}~\bibnamefont {Hagedorn}}\ and\ \bibinfo {author} {\bibfnamefont {A.}~\bibnamefont {Joye}},\ }\href {https://doi.org/10.1007/pl00001017} {\bibfield  {journal} {\bibinfo  {journal} {Ann. Henri Poincar{\'e}}\ }\textbf {\bibinfo {volume} {1}},\ \bibinfo {pages} {837} (\bibinfo {year} {2000})}\BibitemShut {NoStop}%
\bibitem [{\citenamefont {Gradinaru}\ and\ \citenamefont {Hagedorn}(2014)}]{Gradinaru_Hagedorn:2014}%
  \BibitemOpen
  \bibfield  {author} {\bibinfo {author} {\bibfnamefont {V.}~\bibnamefont {Gradinaru}}\ and\ \bibinfo {author} {\bibfnamefont {G.~A.}\ \bibnamefont {Hagedorn}},\ }\href {https://doi.org/10.1007/s00211-013-0560-6} {\bibfield  {journal} {\bibinfo  {journal} {Numer. Math.}\ }\textbf {\bibinfo {volume} {126}},\ \bibinfo {pages} {53} (\bibinfo {year} {2014})}\BibitemShut {NoStop}%
\bibitem [{\citenamefont {Ohsawa}(2018)}]{Ohsawa:2018}%
  \BibitemOpen
  \bibfield  {author} {\bibinfo {author} {\bibfnamefont {T.}~\bibnamefont {Ohsawa}},\ }\href {https://doi.org/10.1088/1361-6544/aaa10c} {\bibfield  {journal} {\bibinfo  {journal} {Nonlinearity}\ }\textbf {\bibinfo {volume} {31}},\ \bibinfo {pages} {1807} (\bibinfo {year} {2018})}\BibitemShut {NoStop}%
\bibitem [{\citenamefont {Begu\v{s}i\'{c}}\ \emph {et~al.}(2019)\citenamefont {Begu\v{s}i\'{c}}, \citenamefont {Cordova},\ and\ \citenamefont {Van{\'{i}}{\v{c}}ek}}]{Begusic_Vanicek:2019}%
  \BibitemOpen
  \bibfield  {author} {\bibinfo {author} {\bibfnamefont {T.}~\bibnamefont {Begu\v{s}i\'{c}}}, \bibinfo {author} {\bibfnamefont {M.}~\bibnamefont {Cordova}},\ and\ \bibinfo {author} {\bibfnamefont {J.}~\bibnamefont {Van{\'{i}}{\v{c}}ek}},\ }\href {https://doi.org/10.1063/1.5090122} {\bibfield  {journal} {\bibinfo  {journal} {J.~Chem.\ Phys.}\ }\textbf {\bibinfo {volume} {150}},\ \bibinfo {pages} {154117} (\bibinfo {year} {2019})}\BibitemShut {NoStop}%
\bibitem [{\citenamefont {Begu{\v{s}}i{\'{c}}}\ and\ \citenamefont {Van{\'{i}}{\v{c}}ek}(2020)}]{Begusic_Vanicek:2020}%
  \BibitemOpen
  \bibfield  {author} {\bibinfo {author} {\bibfnamefont {T.}~\bibnamefont {Begu{\v{s}}i{\'{c}}}}\ and\ \bibinfo {author} {\bibfnamefont {J.}~\bibnamefont {Van{\'{i}}{\v{c}}ek}},\ }\href {https://doi.org/10.1063/5.0013677} {\bibfield  {journal} {\bibinfo  {journal} {J.~Chem.\ Phys.}\ }\textbf {\bibinfo {volume} {153}},\ \bibinfo {pages} {024105} (\bibinfo {year} {2020})}\BibitemShut {NoStop}%
\bibitem [{\citenamefont {van Wilderen}\ \emph {et~al.}(2014)\citenamefont {van Wilderen}, \citenamefont {Messmer},\ and\ \citenamefont {Bredenbeck}}]{VanWilderen_Bredenbeck:2014}%
  \BibitemOpen
  \bibfield  {author} {\bibinfo {author} {\bibfnamefont {L.~J. G.~W.}\ \bibnamefont {van Wilderen}}, \bibinfo {author} {\bibfnamefont {A.~T.}\ \bibnamefont {Messmer}},\ and\ \bibinfo {author} {\bibfnamefont {J.}~\bibnamefont {Bredenbeck}},\ }\href {https://doi.org/10.1002/anie.201305950} {\bibfield  {journal} {\bibinfo  {journal} {Angew. Chem. Int. Ed.}\ }\textbf {\bibinfo {volume} {53}},\ \bibinfo {pages} {2667} (\bibinfo {year} {2014})}\BibitemShut {NoStop}%
\bibitem [{\citenamefont {Yu}\ \emph {et~al.}(2014)\citenamefont {Yu}, \citenamefont {Evans}, \citenamefont {Chatterley}, \citenamefont {Roberts}, \citenamefont {Stavros},\ and\ \citenamefont {Ullrich}}]{Yu_Ullrich:2014}%
  \BibitemOpen
  \bibfield  {author} {\bibinfo {author} {\bibfnamefont {H.}~\bibnamefont {Yu}}, \bibinfo {author} {\bibfnamefont {N.~L.}\ \bibnamefont {Evans}}, \bibinfo {author} {\bibfnamefont {A.~S.}\ \bibnamefont {Chatterley}}, \bibinfo {author} {\bibfnamefont {G.~M.}\ \bibnamefont {Roberts}}, \bibinfo {author} {\bibfnamefont {V.~G.}\ \bibnamefont {Stavros}},\ and\ \bibinfo {author} {\bibfnamefont {S.}~\bibnamefont {Ullrich}},\ }\href {https://doi.org/10.1021/jp507201a} {\bibfield  {journal} {\bibinfo  {journal} {J.~Phys.\ Chem.~A}\ }\textbf {\bibinfo {volume} {118}},\ \bibinfo {pages} {9438} (\bibinfo {year} {2014})}\BibitemShut {NoStop}%
\bibitem [{\citenamefont {Baiardi}\ \emph {et~al.}(2014)\citenamefont {Baiardi}, \citenamefont {Bloino},\ and\ \citenamefont {Barone}}]{Baiardi_Barone:2014}%
  \BibitemOpen
  \bibfield  {author} {\bibinfo {author} {\bibfnamefont {A.}~\bibnamefont {Baiardi}}, \bibinfo {author} {\bibfnamefont {J.}~\bibnamefont {Bloino}},\ and\ \bibinfo {author} {\bibfnamefont {V.}~\bibnamefont {Barone}},\ }\href {https://doi.org/10.1063/1.4895534} {\bibfield  {journal} {\bibinfo  {journal} {J.~Chem.\ Phys.}\ }\textbf {\bibinfo {volume} {141}},\ \bibinfo {pages} {114108} (\bibinfo {year} {2014})}\BibitemShut {NoStop}%
\bibitem [{\citenamefont {Whaley-Mayda}\ \emph {et~al.}(2021)\citenamefont {Whaley-Mayda}, \citenamefont {Guha}, \citenamefont {Penwell},\ and\ \citenamefont {Tokmakoff}}]{WhaleyMayda_Tokmakoff:2021}%
  \BibitemOpen
  \bibfield  {author} {\bibinfo {author} {\bibfnamefont {L.}~\bibnamefont {Whaley-Mayda}}, \bibinfo {author} {\bibfnamefont {A.}~\bibnamefont {Guha}}, \bibinfo {author} {\bibfnamefont {S.~B.}\ \bibnamefont {Penwell}},\ and\ \bibinfo {author} {\bibfnamefont {A.}~\bibnamefont {Tokmakoff}},\ }\href {https://doi.org/10.1021/jacs.1c00542} {\bibfield  {journal} {\bibinfo  {journal} {J.~Am.\ Chem.\ Soc.}\ }\textbf {\bibinfo {volume} {143}},\ \bibinfo {pages} {3060} (\bibinfo {year} {2021})}\BibitemShut {NoStop}%
\bibitem [{\citenamefont {Lau}\ \emph {et~al.}(2023)\citenamefont {Lau}, \citenamefont {DeWitt}, \citenamefont {Boyer}, \citenamefont {Babin}, \citenamefont {Solomis}, \citenamefont {Grellmann}, \citenamefont {Asmis}, \citenamefont {McCoy},\ and\ \citenamefont {Neumark}}]{Lau_Neumark:2023}%
  \BibitemOpen
  \bibfield  {author} {\bibinfo {author} {\bibfnamefont {J.~A.}\ \bibnamefont {Lau}}, \bibinfo {author} {\bibfnamefont {M.}~\bibnamefont {DeWitt}}, \bibinfo {author} {\bibfnamefont {M.~A.}\ \bibnamefont {Boyer}}, \bibinfo {author} {\bibfnamefont {M.~C.}\ \bibnamefont {Babin}}, \bibinfo {author} {\bibfnamefont {T.}~\bibnamefont {Solomis}}, \bibinfo {author} {\bibfnamefont {M.}~\bibnamefont {Grellmann}}, \bibinfo {author} {\bibfnamefont {K.~R.}\ \bibnamefont {Asmis}}, \bibinfo {author} {\bibfnamefont {A.~B.}\ \bibnamefont {McCoy}},\ and\ \bibinfo {author} {\bibfnamefont {D.~M.}\ \bibnamefont {Neumark}},\ }\href {https://doi.org/10.1021/acs.jpca.3c00484} {\bibfield  {journal} {\bibinfo  {journal} {J.~Phys.\ Chem.~A}\ }\textbf {\bibinfo {volume} {127}},\ \bibinfo {pages} {3133} (\bibinfo {year} {2023})}\BibitemShut {NoStop}%
\bibitem [{\citenamefont {Horz}\ \emph {et~al.}(2023)\citenamefont {Horz}, \citenamefont {Masood}, \citenamefont {Brunst}, \citenamefont {Cerezo}, \citenamefont {Picconi}, \citenamefont {Vormann}, \citenamefont {Niraghatam}, \citenamefont {Van~Wilderen}, \citenamefont {Bredenbeck}, \citenamefont {Santoro},\ and\ \citenamefont {Burghardt}}]{Horz_Burghardt:2023}%
  \BibitemOpen
  \bibfield  {author} {\bibinfo {author} {\bibfnamefont {M.}~\bibnamefont {Horz}}, \bibinfo {author} {\bibfnamefont {H.~M.~A.}\ \bibnamefont {Masood}}, \bibinfo {author} {\bibfnamefont {H.}~\bibnamefont {Brunst}}, \bibinfo {author} {\bibfnamefont {J.}~\bibnamefont {Cerezo}}, \bibinfo {author} {\bibfnamefont {D.}~\bibnamefont {Picconi}}, \bibinfo {author} {\bibfnamefont {H.}~\bibnamefont {Vormann}}, \bibinfo {author} {\bibfnamefont {M.~S.}\ \bibnamefont {Niraghatam}}, \bibinfo {author} {\bibfnamefont {L.~J. G.~W.}\ \bibnamefont {Van~Wilderen}}, \bibinfo {author} {\bibfnamefont {J.}~\bibnamefont {Bredenbeck}}, \bibinfo {author} {\bibfnamefont {F.}~\bibnamefont {Santoro}},\ and\ \bibinfo {author} {\bibfnamefont {I.}~\bibnamefont {Burghardt}},\ }\href {https://doi.org/10.1063/5.0132608} {\bibfield  {journal} {\bibinfo  {journal} {J.~Chem.\ Phys.}\ }\textbf {\bibinfo {volume} {158}},\ \bibinfo {pages} {064201} (\bibinfo {year} {2023})}\BibitemShut {NoStop}%
\bibitem [{\citenamefont {Lee}\ and\ \citenamefont {Heller}(1982)}]{Lee_Heller:1982}%
  \BibitemOpen
  \bibfield  {author} {\bibinfo {author} {\bibfnamefont {S.-Y.}\ \bibnamefont {Lee}}\ and\ \bibinfo {author} {\bibfnamefont {E.~J.}\ \bibnamefont {Heller}},\ }\href {https://doi.org/10.1063/1.443342} {\bibfield  {journal} {\bibinfo  {journal} {J.~Chem.\ Phys.}\ }\textbf {\bibinfo {volume} {76}},\ \bibinfo {pages} {3035} (\bibinfo {year} {1982})}\BibitemShut {NoStop}%
\bibitem [{\citenamefont {Tannor}\ and\ \citenamefont {Heller}(1982)}]{Tannor_Heller:1982}%
  \BibitemOpen
  \bibfield  {author} {\bibinfo {author} {\bibfnamefont {D.~J.}\ \bibnamefont {Tannor}}\ and\ \bibinfo {author} {\bibfnamefont {E.~J.}\ \bibnamefont {Heller}},\ }\href {https://doi.org/10.1063/1.443643} {\bibfield  {journal} {\bibinfo  {journal} {J.~Chem.\ Phys.}\ }\textbf {\bibinfo {volume} {77}},\ \bibinfo {pages} {202} (\bibinfo {year} {1982})}\BibitemShut {NoStop}%
\bibitem [{\citenamefont {Baiardi}\ \emph {et~al.}(2013)\citenamefont {Baiardi}, \citenamefont {Bloino},\ and\ \citenamefont {Barone}}]{Baiardi_Barone:2013}%
  \BibitemOpen
  \bibfield  {author} {\bibinfo {author} {\bibfnamefont {A.}~\bibnamefont {Baiardi}}, \bibinfo {author} {\bibfnamefont {J.}~\bibnamefont {Bloino}},\ and\ \bibinfo {author} {\bibfnamefont {V.}~\bibnamefont {Barone}},\ }\href {https://doi.org/10.1021/ct400450k} {\bibfield  {journal} {\bibinfo  {journal} {J.~Chem.\ Theory Comput.}\ }\textbf {\bibinfo {volume} {9}},\ \bibinfo {pages} {4097} (\bibinfo {year} {2013})}\BibitemShut {NoStop}%
\bibitem [{\citenamefont {Bonfanti}\ \emph {et~al.}(2018)\citenamefont {Bonfanti}, \citenamefont {Petersen}, \citenamefont {Eisenbrandt}, \citenamefont {Burghardt},\ and\ \citenamefont {Pollak}}]{Bonfanti_Pollak:2018}%
  \BibitemOpen
  \bibfield  {author} {\bibinfo {author} {\bibfnamefont {M.}~\bibnamefont {Bonfanti}}, \bibinfo {author} {\bibfnamefont {J.}~\bibnamefont {Petersen}}, \bibinfo {author} {\bibfnamefont {P.}~\bibnamefont {Eisenbrandt}}, \bibinfo {author} {\bibfnamefont {I.}~\bibnamefont {Burghardt}},\ and\ \bibinfo {author} {\bibfnamefont {E.}~\bibnamefont {Pollak}},\ }\href {https://doi.org/10.1021/acs.jctc.8b00355} {\bibfield  {journal} {\bibinfo  {journal} {J.~Chem.\ Theory Comput.}\ }\textbf {\bibinfo {volume} {14}},\ \bibinfo {pages} {5310} (\bibinfo {year} {2018})}\BibitemShut {NoStop}%
\bibitem [{\citenamefont {Patoz}\ \emph {et~al.}(2018)\citenamefont {Patoz}, \citenamefont {Begu\v{s}i{\'{c}}},\ and\ \citenamefont {Van{\'{i}}{\v{c}}ek}}]{Patoz_Vanicek:2018}%
  \BibitemOpen
  \bibfield  {author} {\bibinfo {author} {\bibfnamefont {A.}~\bibnamefont {Patoz}}, \bibinfo {author} {\bibfnamefont {T.}~\bibnamefont {Begu\v{s}i{\'{c}}}},\ and\ \bibinfo {author} {\bibfnamefont {J.}~\bibnamefont {Van{\'{i}}{\v{c}}ek}},\ }\href {https://doi.org/10.1021/acs.jpclett.8b00827} {\bibfield  {journal} {\bibinfo  {journal} {J.~Phys.\ Chem.\ Lett.}\ }\textbf {\bibinfo {volume} {9}},\ \bibinfo {pages} {2367} (\bibinfo {year} {2018})}\BibitemShut {NoStop}%
\bibitem [{\citenamefont {Kundu}\ \emph {et~al.}(2022)\citenamefont {Kundu}, \citenamefont {Roy}, \citenamefont {Fleming},\ and\ \citenamefont {Makri}}]{Kundu_Makri:2022}%
  \BibitemOpen
  \bibfield  {author} {\bibinfo {author} {\bibfnamefont {S.}~\bibnamefont {Kundu}}, \bibinfo {author} {\bibfnamefont {P.~P.}\ \bibnamefont {Roy}}, \bibinfo {author} {\bibfnamefont {G.~R.}\ \bibnamefont {Fleming}},\ and\ \bibinfo {author} {\bibfnamefont {N.}~\bibnamefont {Makri}},\ }\href {https://doi.org/10.1021/acs.jpcb.2c00846} {\bibfield  {journal} {\bibinfo  {journal} {J.~Phys.\ Chem.~B}\ }\textbf {\bibinfo {volume} {126}},\ \bibinfo {pages} {2899} (\bibinfo {year} {2022})}\BibitemShut {NoStop}%
\bibitem [{\citenamefont {Wenzel}\ and\ \citenamefont {Mitric}(2023)}]{Wenzel_Mitric:2023}%
  \BibitemOpen
  \bibfield  {author} {\bibinfo {author} {\bibfnamefont {M.}~\bibnamefont {Wenzel}}\ and\ \bibinfo {author} {\bibfnamefont {R.}~\bibnamefont {Mitric}},\ }\href {https://doi.org/10.1063/5.0130340} {\bibfield  {journal} {\bibinfo  {journal} {J.~Chem.\ Phys.}\ }\textbf {\bibinfo {volume} {158}},\ \bibinfo {pages} {034105} (\bibinfo {year} {2023})}\BibitemShut {NoStop}%
\end{thebibliography}
\clearpage % Ensure it's on a new page
\onecolumngrid
\pagestyle{empty}
{
 \renewcommand{\newpage}{\par\pagebreak}
  \includepdf[pages={-},angle=0,offset=0 0,noautoscale=true]{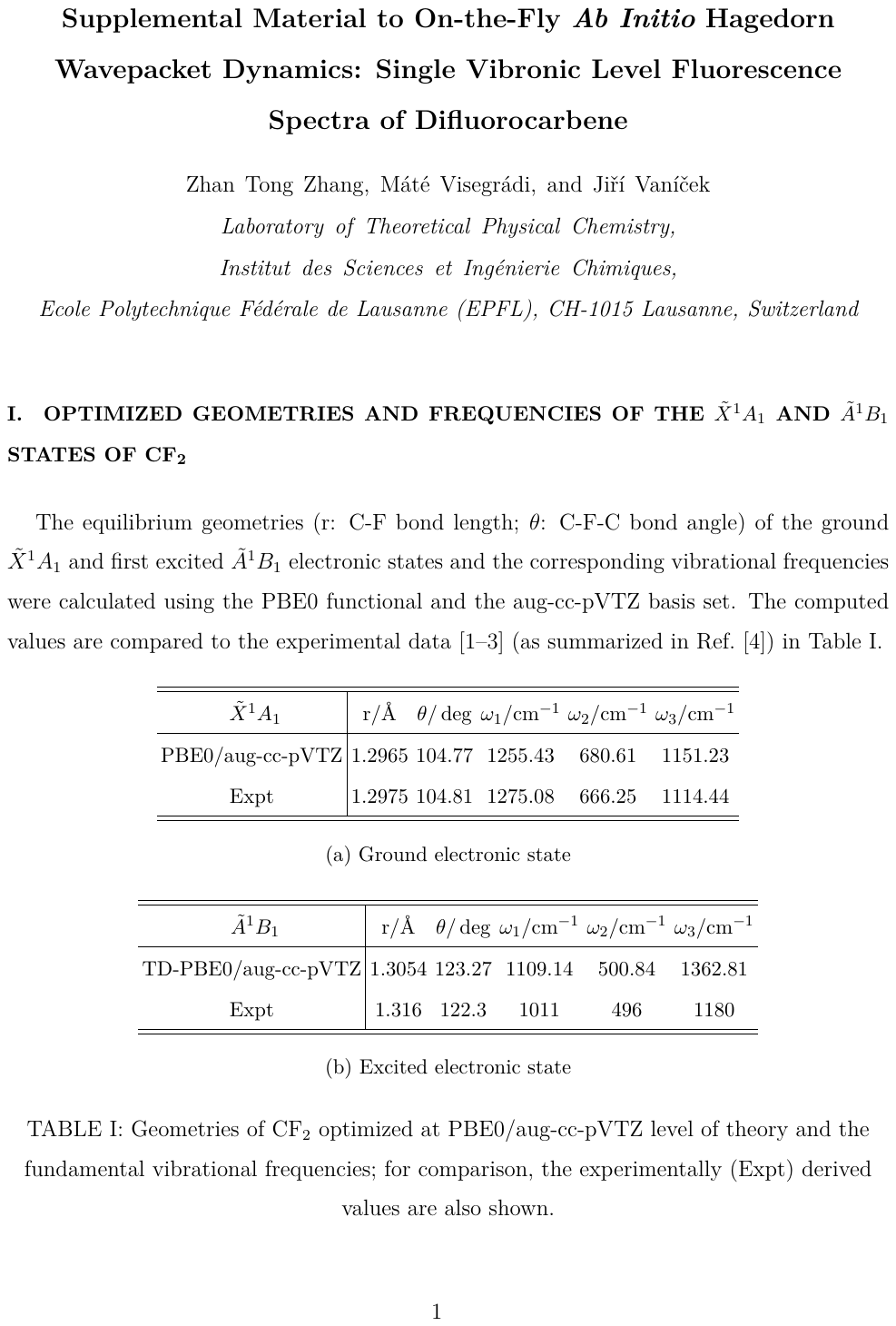}
}
\end{document}